\documentclass[
]{article}

\usepackage{arxiv}

\usepackage[utf8]{inputenc} % allow utf-8 input
\usepackage[T1]{fontenc}    % use 8-bit T1 fonts

\usepackage{booktabs}       % professional-quality tables
\usepackage{amsfonts}       % blackboard math symbols
\usepackage{nicefrac}       % compact symbols for 1/2, etc.
\usepackage{microtype}      % microtypography
\usepackage{lipsum}		% Can be removed after putting your text content
\usepackage[
	]{graphicx}
\graphicspath{{figures/}} % Set the default folder for images
\usepackage{doi}
\usepackage{url}            % simple URL typesetting

\usepackage[
	style=authoryear
	,firstinits
	,url=false
	,natbib=true
	,backend=biber
	,maxcitenames=2
	,citestyle=numeric-comp
	]{biblatex}
\renewbibmacro{in:}{}
\DeclareFieldFormat*{title}{#1}
\bibliography{references.bib}

\usepackage[dvipsnames]{xcolor}
\usepackage{amsmath, amssymb}
\usepackage{amsthm}
\usepackage{euler}
\usepackage{siunitx}
\usepackage{xfrac}

\DeclareMathAlphabet{\altmathcal}{OMS}{cmsy}{m}{n}
\definecolor{link-color}{HTML}{548687}
\definecolor{cite-color}{HTML}{B0413E}

\title{Lagged teleconnections of climate variables\\ identified via complex rotated Maximum Covariance Analysis}

\author{ \includegraphics[scale=0.06]{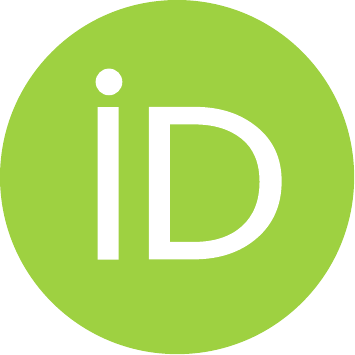}\hspace{1mm}\href{https://orcid.org/0000-0003-3357-1742}{Niclas Rieger} \\
	Centre de Recerca Matem\`atica (CRM)\\
	Departament de F\'{\i}sica\\
	Universitat Aut\`onoma de Barcelona\\
	Bellaterra, Spain\\
	\href{mailto:nrieger@crm.cat}{\texttt{nrieger@crm.cat}} \\
	\And
	\includegraphics[scale=0.06]{orcid.pdf}\hspace{1mm}\href{https://orcid.org/0000-0002-5280-2692}{\'Alvaro Corral} \\
	Centre de Recerca Matem\`atica (CRM)\\
	Bellaterra, Spain\\
	Complexity Science Hub Vienna\\
	Vienna, Austria\\
	\href{mailto:acorral@crm.cat}{\texttt{acorral@crm.cat}}
	\AND
	Estrella Olmedo \\
	Institute of Marine Sciences (ICM-CSIC)\\
	Barcelona Expert Center (BEC)\\
	Barcelona, Spain \\
	\href{mailto:olmedo@icm.csic.es}{\texttt{olmedo@icm.csic.es}}
	\And
	\includegraphics[scale=0.06]{orcid.pdf}\hspace{1mm}\href{https://orcid.org/0000-0001-6103-224X}{Antonio Turiel} \\
	Institute of Marine Sciences (ICM-CSIC)\\
	Barcelona Expert Center (BEC)\\
	Barcelona, Spain \\
	\href{mailto:turiel@icm.csic.es}{\texttt{turiel@icm.csic.es}}
}

\renewcommand{\headeright}{Preprint}
\renewcommand{\undertitle}{\color{gray} \small This work has been submitted to the Journal of Climate. This work has not yet been peer-reviewed and is provided by
the contributing author(s) as a means to ensure timely dissemination of scholarly and technical work on a
noncommercial basis. Copyright in this work may be transferred without further notice.}
\renewcommand{\shorttitle}{}

\hypersetup{
	colorlinks=true
	,breaklinks=true
	,bookmarks=true
	,bookmarksnumbered
	,urlcolor=black
	,linkcolor=cite-color
	,citecolor=cite-color % Link colors
	,pdftitle={Lagged teleconnections of climate variables identified via complex rotated Maximum Covariance Analysis}
	,pdfsubject={physics.ao-ph}
	,pdfauthor={Niclas Rieger, Alvaro Corral, Estrella Olmedo, Antonio Turiel}
	,pdfkeywords={MCA, Varimax, Promax, rotation, Hilbert, teleconnections}
}

\begin{document}
\maketitle

\begin{abstract}
	A proper description of ocean-atmosphere interactions is key for a correct understanding of climate evolution. The interplay among the different variables acting over the climate is complex, often  leading to correlations across long spatial distances (teleconnections). In some occasions, those teleconnections occur with quite significant temporal shifts that are fundamental for the understanding of the underlying phenomena but which are poorly captured by standard methods. Applying orthogonal decomposition such as Maximum Covariance Analysis (MCA) to geophysical data sets allows to extract common dominant patterns between two different variables, but generally suffers from (i) the non-physical orthogonal constraint as well as (ii) the consideration of simple correlations, whereby temporally offset signals are not detected. Here we propose an extension, complex rotated MCA, to address both limitations. We transform our signals using the Hilbert transform and perform the orthogonal decomposition in complex space, allowing us to correctly correlate out-of-phase signals. Subsequent Varimax rotation removes the orthogonal constraints, leading to more physically meaningful modes of geophysical variability. As an example of application, we have employed this method on sea surface temperature and continental precipitation; our method successfully captures the temporal and spatial interactions between these two variables, namely for (i) the seasonal cycle, (ii) canonical ENSO, (iii) the global warming trend, (iv) the Pacific Decadal Oscillation, (v) ENSO Modoki and finally (vi) the Atlantic Meridional Mode. The complex rotated modes of MCA provide information on the regional amplitude, and under certain conditions, the regional time lag between changes on ocean temperature and  land precipitation.
\end{abstract}

% keywords can be removed
\keywords{MCA \and rotation \and complex \and Hilbert transform \and lagged teleconnections \and xmca}

\section*{Significance Statement}
Correlations between time series of different climate variables are often time-lagged and can appear over long spatial distances. Our goal was to develop a method that allows the simultaneous identification of the dominant spatial patterns and their time lags between two different climate variables. Using sea surface temperatures and continental rainfalls as example, our method extracts the different dynamics of the seasonal cycle as well as 5 other well-known climate phenomena. Especially for cyclic time series like the seasonal cycle, the relative time lags at different locations can be determined precisely, whereas for acyclic time series only qualitative statements about time lags can be made. In future studies, we expect new insights into the dynamical structure of the Madden-Julian oscillation thanks to this method, which is readily available as a Python package.

\section{Introduction}\label{sec:intro}

The Earth's climate system is extremely complicated and deciphering the web of interdependencies and influences of different climate subsystems is an involved challenge. As the quantity and quality of Earth observations increase thanks to the advances in remote sensing, so too does the amount of data that needs to be processed. Data-driven dimensionality reduction methods are therefore crucial for climate studies, as they allow high-dimensional spatio-temporally resolved signals to be disaggregated into the dominant patterns, while still capturing the subtle details of higher resolution data. As such, Principal Component Analysis (PCA), or Empirical Orthogonal Functions (EOF) analysis as it is often referred to in climate science, allows to identify the dominant internal structure of the variability as expressed by the variance, with a variety of different available versions of PCA proving the popularity of such methods in climate science \citep[e.g.][]{hannachi_empirical_2007,hannachi_patterns_2021}.

Climate phenomena with different expression in oceanic and atmospheric variables, such as the El Niño-Southern Oscillation (ENSO), however, require the simultaneous analysis of several variables for a more comprehensive description. In principle, multivariate PCA \citep{kutzbach_empirical_1967} makes it possible to extract the patterns of co-variability of more than one variable. However, multivariate PCA accumulates the variance and the covariance of variables with very different variability in the same quantities. In consequence this may mask co-varying patterns as low-variability patterns of one variable can be erroneously accumulated in very dominant structures of one of the other, large-variability variables \citep{bretherton_intercomparison_1992}.

Maximum Covariance Analysis (MCA)\footnote{Sometimes referred to as Singular Value Decomposition (SVD) analysis. This name is unfortunate and should not be confused with the actual factorisation technique of a real/complex matrix.} avoids this masking by taking into account only the covariance between two sets of variables. As such, it bears similarity to Canonical Correlation Analysis \citep[CCA,][]{hotelling_relations_1936} which aims at maximising the temporal correlation between both variables. When the number of grid points (i.e. number of time series) is higher than the number of observations (i.e. number of time steps)
and the data exhibits multicollinearity, as it is often the case for climate data, CCA fails as it requires the individual variance matrices to be non-singular unless regularised \citep{vinod_canonical_1976,cruz-cano_fast_2014,hannachi_regularised_2016}. In case the two fields of variables are identical, MCA reduces to PCA, the former thus being a natural generalisation of PCA.

Yet the methods discussed above maximise instantaneous correlation and do not consider time-delayed signals. To gain a deeper understanding of the dynamics of climate phenomena, however, it is necessary to systematically investigate time lags. A typical approach to tackle with this problem is to consider one variable set with a time lag defined \textit{a priori} followed by a MCA \citep[e.g.][]{li_north_2016}. However, this requires knowledge of the time lag which may vary from one location to another \citep{ballabrerapoy_potential_2002}.

In this paper, we propose \textit{complex rotated MCA} to systematically investigate the phase shift of two variables. We generate complex time series known as the analytical signal, where the real and imaginary parts are related to each other by the Hilbert transform, and decompose the covariance matrix in complex space, in analogy to complex PCA \citep{horel_complex_1984,bloomfield_orthogonal_1994}. We also effectively reduce spectral leakage inherent in the Hilbert transform of non-cyclic signals by using an extrapolation method of the signal beyond its boundaries. Finally, to relax the orthogonality constraint of the obtained solutions, we apply Varimax rotation to the spatial patterns, which leads to more localised solutions and thus facilitates their physical interpretation \citep{richman_rotation_1986,cheng_orthogonal_1995}.

To make the method readily accessible as a tool, we provide it as a Python package, called \textit{xmca} \citep{rieger_xmca_2021}. Due to the power and popularity of NumPy \citep{van_der_walt_numpy_2011} and xarray \citep{hoyer_xarray_2017}, both packages form the basis of xmca, so that their typical data format can be used directly as input for analysis. The package is modularised in a way that provides the user free choice whether standard, complex, rotated or complex rotated MCA is to be performed. The user can also choose between Varimax orthogonal rotation as well as Promax oblique rotation. Further, if desired, standardisation of the input data is computed on the fly. The different flavours work in the same way for PCA, if one instead of two fields is provided as input.

The remainder of the article is structured as follows. Section \ref{sec:methods} introduces the methodology, where we briefly discuss MCA (Sec.~\ref{sec:mca}), complex MCA (Sec.~\ref{sec:cmca}) and rotated MCA (Sec.~\ref{sec:rmca}). Section~\ref{sec:data} describes the data used to test the method using first synthetic data (Sec.~\ref{sec:data_synthetic}) and then climatic variables (Sec.~\ref{sec:data_climate}).  Section~\ref{sec:results} presents the results of both the synthetic (Sec.~\ref{sec:results_synthetic}) and real-world analysis (Sec.~\ref{sec:results_climate}). We conclude our study and provide directions for future research in Section~\ref{sec:conclusion}.

\section{Methods}\label{sec:methods}

\subsection{Maximum Covariance Analysis}\label{sec:mca}
Let us consider two spatio-temporal data fields $\pmb{X}_A \in \mathbb{R}^{m \times n_A}$ and $\pmb{X}_B \in \mathbb{R}^{m \times n_B}$ representing two different geophysical fields
$s \in \left\{A,B\right\}$, both having temporal dimension $m$ and spatial dimensions $n_A$ and $n_B$, respectively. Throughout the text, the index $s$ is used implicitly without further definition to represent one of the two fields. In the following, we will refer to the temporal dimensions as the number of observations while we denote the spatial dimensions by the number of grid points. Assuming each time series to have zero mean, MCA then aims at maximising
\begin{align}
    \pmb{v}_A^T \pmb{C} \pmb{v}_B, \quad s.t. \quad \pmb{v}_A^T\pmb{v}_A = \pmb{v}_B^T\pmb{v}_B = 1
\end{align}
where $\pmb{C}$ denotes the temporal covariance matrix and $\pmb{v}_A,\pmb{v}_B$ the \textit{spatial patterns}, of both fields, respectively. Mathematically, this can be achieved by applying the singular value decomposition (SVD) to the covariance matrix,
\begin{align}
    \pmb{C} = \frac{1}{m-1}\pmb{X}_A^T \pmb{X}_B = \pmb{V}_A \pmb{\Sigma} \pmb{V}_B^T,
\end{align}
with the columns of the obtained singular vector matrices $(\pmb{V}_s)$ representing pairs of spatial patterns describing the maximum amount of temporal covariance between both variables. The entries of $\pmb{\Sigma} \in \mathbb{R}^{n_A \times n_B}$ along the main diagonal, the singular values $\sigma_k$, represent the covariance of each spatial pattern pair $k$, providing a mean of estimating the relative importance of each pair via the covariance fraction\footnote{Typically the squared covariance fraction defined as $\gamma_k^* = \sigma_k^2 / \left( \sum_{j=1}^{\min(n_A,n_B)} \sigma_j^2 \right)$ is considered for the relative importance of each mode for MCA. However, we opt for the non-squared covariance fraction since the total explained covariance is conserved under rotation, i.e. for $r$ rotated modes $\sum_i^r \sigma_i = \sum_i^r \sigma_i^*$ where $\sigma_i^*$ refers to the covariance associated to mode $i$ after rotation. Furthermore, this measure is comparable to the solutions obtained by PCA, and in fact it is equivalent when $\pmb{X}_A=\pmb{X}_B$, for which MCA reduces to PCA and the singular values equal the eigenvalues in PCA.} $\gamma_k$
\begin{align}\label{eq:cf}
    \gamma_k = \sigma_k  \left( \sum_{j=1}^{\min(n_A,n_B)} \sigma_j \right)^{-1}.
\end{align}

By projecting the data fields on their respective singular vectors, we obtain the corresponding temporal evolution for each spatial pattern given by the columns of $\pmb{P}_s = \pmb{X}_s\pmb{V}_s$. Since the singular vectors are orthonormal, i.e. $\pmb{V}_s^T\pmb{V}_s = \pmb{1}_{n_s}$ with $\pmb{1}_{n_s}$ being the identity matrix of rank $n_s$, the projections of the left and right field are uncorrelated, i.e. $\pmb{P}_A^T \pmb{P}_B = \pmb{\Sigma} / (m-1)$, while the projections of the same field are correlated in general, i.e. $\pmb{P}_s^T \pmb{P}_s$ is not a diagonal matrix. In this paper, we will refer to the spatial patterns and their corresponding time projections as \textit{empirical orthogonal functions} (EOFs) and \textit{principal components} (PCs), respectively, according to the usual convention in climate science. The EOFs and the PCs associated with a specific singular value $\sigma_k$ are denoted as \textit{mode $k$}.

\subsection{Complex MCA}\label{sec:cmca}
Propagating features or lagged signals could be detected by using a complex representation of the input fields. In analogy to complex PCA \citep{wallace_empirical_1972, rasmusson_biennial_1981, horel_complex_1984, bloomfield_orthogonal_1994}, we complexify the real input fields via the Hilbert transform to construct the \textit{analytical signal} $\hat{\pmb{X}}_s$ defined as
\begin{align}
    \hat{\pmb{X}}_s = \pmb{X}_s + i \altmathcal{H}(\pmb{X}_s)
\end{align}
where $\altmathcal{H}(\cdot)$ denotes the column-wise applied Hilbert transform. The analytical signal constructed in that way is a unique complex representation of the real signal, but whether it also represents a physical reality depends on the frequency spectrum of the analysed signal. By construction, the frequency components of the Hilbert transform are phase shifted by $-\sfrac{\pi}{2}$ with respect to those of the original signal. Therefore, for narrow-bandwidth signals, the Hilbert transform has a simple physical interpretation i.e., it represents a signal which arrives with a lag of one fourth of the typical period. If the signal consists of multiple dominating frequencies, however, the interpretation of the phase is more elusive, as it  cannot be simply associated with a single frequency. Thus, the more narrow the signal bandwidth, the more directly we can relate the phase to specific timings of its Hilbert transform \citep{boashash_estimating_1992}.

\begin{figure*}[t]
    \noindent\includegraphics[width=\textwidth,angle=0]{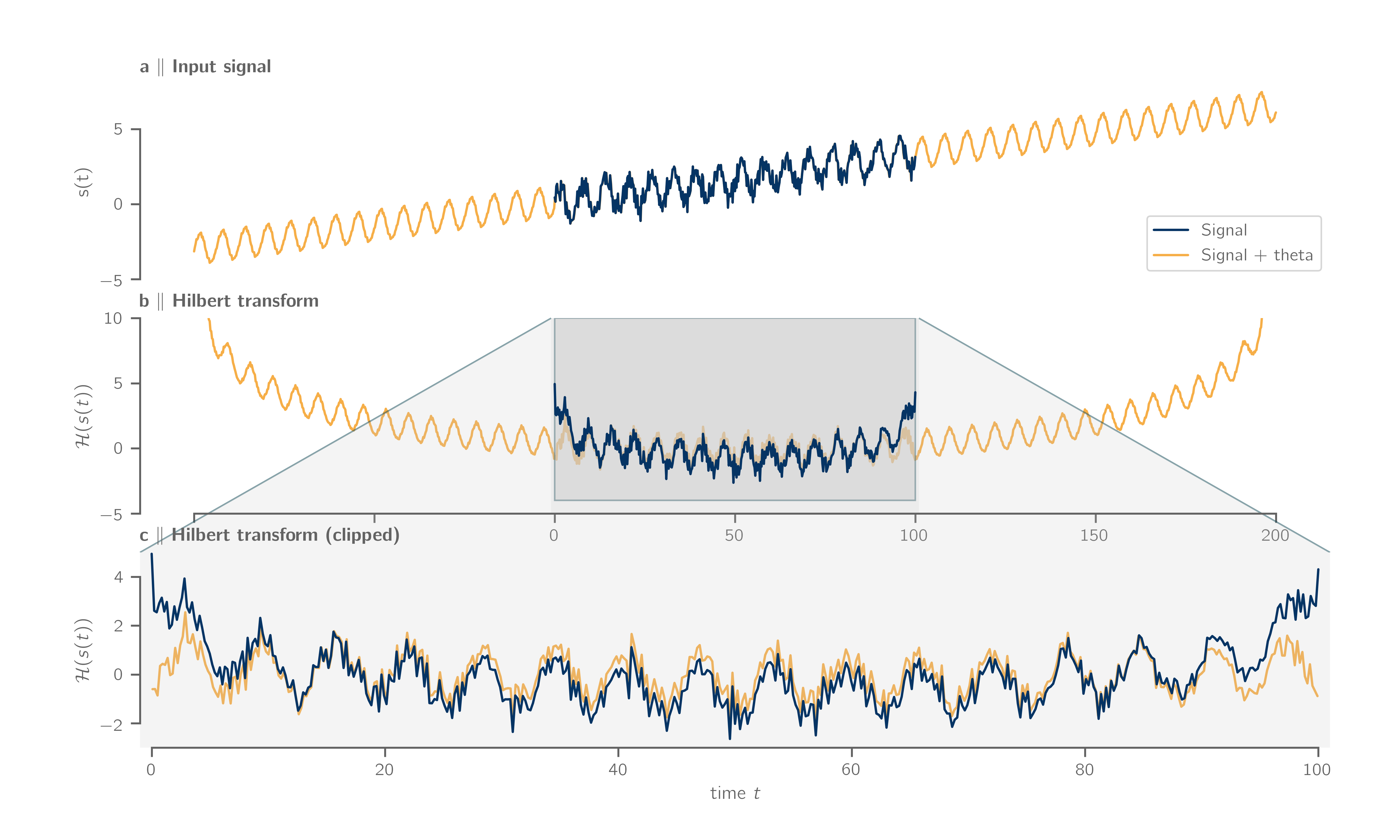}
    \caption{Example illustrating the Hilbert transform using the Theta extension. \textbf{(a)} Input signal $s(t)=sin(t) + ct + \epsilon(t)$ using an arbitrary constant $c$ and Gaussian white noise $\epsilon$ with zero mean and unit variance (blue) as well as the extended time series via forecasting/backcasting using the optimised Theta model \citep{fiorucci_models_2016}. \textbf{(b)} Hilbert transform $\altmathcal{H(\cdot)}$ of the original and extended signal, respectively. \textbf{(c)} Comparison of both Hilbert transforms over the domain of the original series.}
    \label{fig:example_hilbert_theta}
\end{figure*}

A fundamental issue in the computation of the Hilbert transform arises when non-stationary or drifting signals are processed. Such signals are non-cyclic, and therefore, when the Fourier coefficients are calculated, strong boundary effects can occur due to spectral leakage \citep{boashash_estimating_1992} (Fig.~\ref{fig:example_hilbert_theta}). This problem can be circumvented by detrending the time series and considering only integer cycles as well as by applying window functions to the time series (e.g. Hanning, Hamming). However, this comes at the cost of information loss. Additionally, such mitigation techniques are particularly ill suited to deal with non-stationarities associated to climate change (that include not only noticeable trends on the mean level, but also increases in the amplitude of some periodic phenomena). Therefore, it seems important to introduce techniques capable to dealing with non-stationary signals.

To mitigate spectral leakage across the boundaries of the time series, we extrapolate the time series at both boundaries, to the past and to the future, using the optimised Theta model \citep{assimakopoulos_theta_2000,fiorucci_models_2016}, a special case of an autoregressive integrated moving average model with drift, ARIMA(0,1,1) \citep{hyndman_unmasking_2003}. The Theta model is a relatively simple yet well performing extrapolation method. While the forecast in itself actually works with non-cyclic signals, seasonal features are considered via multiplicative classical decomposition, thus allowing cyclic and non-cyclic signals to be extrapolated. This approach of handling the seasonal structure of a time series requires the user to specify the dominant period of the signal, $T_s$, beforehand. For discrete time series $T_s$ represents the number of time steps needed to complete one cycle, that is, e.g. $365$ for daily data considering an annual cycle, or $24$ for hourly data with a daily cycle (for more details we refer the reader to \citet{fiorucci_models_2016}). We then apply the Hilbert transform to the extended series, so the spectral leakage is only important on the backward and forward extensions of it. Finally, we extract the central part (removing the parts corresponding to the extension), that correctly corresponds to the Hilbert transform of the original series.  Using this approach, we effectively reduce the edge effects of the Hilbert transform compared to a non-processed time series (Fig.~\ref{fig:example_hilbert_theta}). Notice that it is not necessary that the extrapolation faithfully reproduces the characteristics of the original series; it just suffices for our purpose that the extrapolated time series approximately continues the cyclic structure at the original time series boundaries in order to reduce spectral leakage.  Apart from tracking trends, the exact extrapolation beyond the boundaries is not essential since its effects on the central part of the Hilbert transform are very marginal at most.

After the described complexification of the original time series, we follow the steps of standard MCA, with the difference that the transpose $(\cdot)^T$ incorporates the complex conjugate $\bar{(\cdot)}$ and the obtained EOFs and projections PCs are complex and unitary. This allows us to calculate the spatial amplitude $\pmb{\mathcal{A}}_s$ and phase function $\pmb{\Theta}_s$ for both fields,
\begin{align}\label{eq:amplitude}
    \pmb{\mathcal{A}}_{s} =& \left( \pmb{V}_{s} \odot \pmb{\bar{V}}_{s} \right)^{\odot \frac{1}{2}}, \\ \label{eq:phase}
    \pmb{\Theta}_{s} =& \arctan\!2 \left( \text{Im} (\pmb{V}_{s}), \text{Re}~ (\pmb{V}_{s}) \right),
\end{align}
where $\odot$ denotes the element-wise multiplication/exponentiation, and $\arctan\!2$ refers to the two-argument arctangent which is calculated element-wise (Appendix~\ref{app:atan2}). Although this matrix notation seems somewhat cumbersome compared to the more direct expression through scalar fields, it allows us to be coherent with the rest of the paper. The phase function can be interpreted directly as a time lag if the corresponding (real) PC has a narrow-band spectrum with just one dominant frequency. If the spectrum is rather broad-band or has several dominant frequencies, an interpretation of the phase function is usually not straightforward. We note that the complex EOFs derived from the SVD are only defined up to a phase shift of $\exp{i\theta}$ with $\theta \in [0, 2\pi]$. However, if complex EOFs are obtained with a different phase shift $\theta$ (e.g. due to another SVD algorithm), the change will also be reflected in the projected PCs, so that taking into account both, PC and phase function, the results are unambiguous.

\subsection{Rotated MCA}\label{sec:rmca}
While orthogonality is often a mathematically desirable property, it does not make a lot of sense from a purely geophysical standpoint. Therefore, standard EOFs are difficult to interpret in the case of geophysical data. The major drawbacks of EOFs due to orthogonality are twofold: First, EOFs are sensitive to the selected spatial domain, that is including or removing some regions may change large parts of the EOFs. Secondly, EOFs tend to split certain geophysically meaningful patterns across several consecutive modes \citep{richman_rotation_1986}.  To relax the orthogonality constraint to better accommodate the geophysical reality, the EOFs can be \textit{rotated}, which implies a linear transformation of the first $r$ loaded\footnote{Loaded EOFs are weighted by the square root of the corresponding singular value.} EOFs $\pmb{L}_{s,r}$. This concept, which was originally developed in the context of PCA, can also be applied to MCA \citep{cheng_orthogonal_1995}. For this, we apply the rotation matrix $\pmb{R} \in \mathbb{C}^{r \times r}$ to the \textit{loading matrix} $\pmb{L}_r \in \mathbb{C}^{n \times r}$, $n = n_A + n_B$,
\begin{align}\label{eq:loading_matrix}
    \pmb{L}_{r} =& \begin{pmatrix}
    \pmb{L}_{A,r} \\ \pmb{L}_{B,r}
    \end{pmatrix}  = \begin{pmatrix}
    \pmb{V}_{A,r} \\ \pmb{V}_{B,r}
    \end{pmatrix} \pmb{\Sigma}_r^{1/2},
\end{align}
where $r$ reflects the respective submatrices containing only the first $r$ columns. In addition, $\pmb{\Sigma}_r$ is the diagonal submatrix containing only the first $r$ columns and rows.

There are a number of different criteria for defining the rotation matrix $\pmb{R}$ \citep[e.g.][]{richman_rotation_1986}, including the Varimax orthogonal rotation \citep{kaiser_varimax_1958} and the Promax oblique rotation \citep{hendrickson_promax_1964}, whose general aim is to regroup the obtained patterns by approximating simple structures \citep{thurstone_simple_1947}. Mathematically, Varimax rotation seeks to maximise the summed variances of squared loadings which is achieved by (i) restricting rotated EOFs to be composed by only a few numbers of grid points with high loadings while the remaining grid points exhibit near-zero loadings and by (ii) limiting each grid point to contribute to only one rotated EOF while having near-zero loadings for the other EOFs. Since non-rotated EOFs are typically dense, that is consisting of mostly non-zero values, Varimax rotation produces more sparse EOFs containing mostly zero or close-to-zero values, leading to spatially compact structures which allow a clearer interpretation. Promax oblique rotation builds upon the Varimax solution by raising the rotated, normalised EOFs to the power $p\geq1$ while retaining the original sign, thus further reducing low loading compared to high loading of the EOFs. Promax can be understood as an oblique generalisation, with $p=1$ yielding a Varimax orthogonal solution. \citet{hendrickson_promax_1964} provides a value for $p$, which the authors consider appropriate for most applications ($p=4$). In the extensive review \citet{richman_rotation_1986} points out, however, that the Promax rotation using $p=2$ consistently performs better, which is what we will use in this paper. In order to keep the paper self-contained, we provide a brief summary of both rotation criteria in Appendix~\ref{sec:rotation_criterion}.

The main difference between both rotation types is that Promax allows rotated PCs to be correlated with each other, with higher values of $p$ typically leading to stronger correlations. In contrast, Varimax solutions yield always uncorrelated PCs. For both, Varimax orthogonal and Promax oblique rotation, the obtained EOFs are no longer orthogonal.
The question which rotation method is the most suitable for a given analysis remains unsettled in the literature. In reality, we do not expect geophysical signals to be perfectly uncorrelated, which generally argues for applying an oblique rotation. Nevertheless, \citet{finch_comparison_2006} showed that Varimax orthogonal and Promax oblique solutions perform similarly, in particular when the PCs obtained by the oblique solution exhibit low linear Pearson correlation coefficients. In the presence of simple structures, however, Promax oblique rotation performs better by effectively reducing the number of grid points that contribute to each mode, hence further simplifying the EOFs and increasing correlations among the PCs \citep{finch_comparison_2006}. Therefore, the decision on how many EOFs to rotate and which rotation type to perform is a choice to be taken case-by-case and which we will explore in Sec.~\ref{sec:results_climate}.

\section{Data}\label{sec:data}
To test our method, we apply it to artificial and real climate data. For the artificial data sets, we consider complex MCA without rotation, as studies already exist that demonstrate the better interpretability and lower sensitivity to sampling errors of the Varimax-rotated solutions compared to the unrotated EOFs. \citep{lian_evaluation_2012,richman_rotation_1986, cheng_robustness_1995}. By means of two synthetic experiments we seek to illustrate the advantages and caveats of complex MCA. In a first experiment (Experiment I), we test the performance of complex MCA compared to standard MCA considering time-lagged signals. In a second experiment (Experiment II), we investigate how the Theta extension can improve the result of complex MCA to non-stationary processes. Finally, we apply complex MCA with rotation to climatic data that we expect to have intrinsic geophysical cycles but are also affected by the non-stationarity of climate change.

\subsection{Synthetic Data}\label{sec:data_synthetic}
We create two 2D spatio-temporal data fields $X_A,X_B$ with coordinates representing longitude $\lambda \in [0, 359]$ and time $t \in [0, 364]$ days. The data generation model follows
\begin{align}
    X_s(t, \lambda) =& c_s(\lambda) \zeta_s(t,\lambda) + \epsilon(t,\lambda)
\end{align}
where $\zeta_s(t,\lambda)$ depends on the individual experiment design, $c_s(\lambda) = \cos^2{a_s \lambda}$ represents a "zonal" modulation factor with scale factor $a_s$ and $\epsilon(t,\lambda)$ denotes Gaussian white noise with zero mean and variance of $1$. \\
The idea of Experiment I is to highlight the advantage of using complexified fields compared to standard MCA in the presence of moving patterns and phase-shifted, stationary fields. Therefore, we define both signals as travelling waves $\zeta_s(t,\lambda) = \cos{(\omega t + k_s\lambda + \phi_s)}$. Our parameter choices are motivated by the Madden-Julian Oscillation (MJO), which is an eastwards propagating mode of deep convection and associated zonal wind circulation in the tropical atmosphere \citep{madden_detection_1971,zhang_madden-julian_2005}. As such, the MJO is one of the dominant drivers of intraseasonal variability in the tropics characterized by a zonal propagation period of about 30-90 days. In general, MJO events tend to dominate over the Indian Ocean and the western Pacific before they decay towards the eastern part of the Pacific. Furthermore, when observing the MJO through different variables, the spatial scale of MJO events can vary. While the zonal wind circulation typically exhibits a wave number of about $1$, the convective precipitation patterns may have zonal wave numbers of $1-3$. With this in mind, we fix the model parameter to $\omega = \sfrac{2\pi}{56}$ \si{days^{-1}} (representing a period of \SI{56}{days}), $k_A = \sfrac{-2\pi}{360}$, $k_B = \sfrac{-6\pi}{360}$ (representing wave numbers 1 and 3, respectively) and phase shifts $\phi_A = 0$ and $\phi_B = \sfrac{\pi}{2}$. Finally, the modulation factor $c_s(\lambda)$ can be thought of describing regions of enhanced and suppressed MJO activity in the tropics. We therefore choose the scale factor $a_A = \sfrac{\pi}{360}$ and $a_B = \sfrac{2\pi}{360}$ representing 1 and 2 regions of enhanced activity, respectively. (Fig.\ref{fig:toydata1}a). \\
In Experiment II, we investigate the response of complex MCA to non-cyclical, non-stationary signals. Since MCA seeks to maximise covariance through a new set of linear combinations, we restrict ourselves here to linear trends (Tab.~\ref{tab:data}). To keep the experiment as simple as possible, we also consider static fields only. Then, the signal may be simply defined as $\zeta_s(t, \lambda) =  \zeta_s(t) = h_st$, where $h_A = \sfrac{3}{365}$ \si{days^{-1}} and $h_B = \sfrac{5}{365}$ \si{days^{-1}} describe linear trends. In this case, the modulation factor $c_s(\lambda)$ may describe, for instance, differential heating of the continents and the oceans due to global warming. The actual values of the scale factors remain unchanged compared to Experiment I (Fig.~\ref{fig:toydata2}a). For a better overview, the parameters of both experiments are summarised in Tab.~\ref{tab:data}.

\begin{table*}[t]
\caption{Models for synthetic data generation of $X_s$ for Experiment I and II (Section~\ref{sec:data}\ref{sec:data_synthetic}). For our final model, we consider the coordinate ranges $t = 0, 1, \dots, 364$ days and $\lambda= 0,1 \dots, 359$ (wihtout unit).}
\label{tab:data}
\centering
% For LaTeX tables use
\begin{tabular*}{\textwidth}{@{\extracolsep\fill}llllll}
\hline\noalign{\smallskip}
\multicolumn{2}{c}{} &
  \multicolumn{3}{l}{Experiment I} &
  \multicolumn{1}{l}{Experiment II} \\
\hline\noalign{\smallskip}
Field  & $a_s$ &  $\omega$ (\si{days^{-1}}) & $k_s$ & $\phi_s$ &  $h_s$ (\si{days^{-1}})  \\
\noalign{\smallskip}\hline\noalign{\smallskip}
$A$ & $\sfrac{\pi}{360}$ & $\sfrac{2\pi}{56}$ & $-\sfrac{2\pi}{360}$ & $0$ & $\sfrac{3}{365}$ \\
$B$ & $\sfrac{2\pi}{360}$ & $\sfrac{2\pi}{56}$ & $-\sfrac{6\pi}{360}$ & $\sfrac{\pi}{2}$ & $\sfrac{5}{365}$ \\
\noalign{\smallskip}\hline
\end{tabular*}
\end{table*}

\subsection{Climate Data}\label{sec:data_climate}
We analyse monthly means of global sea surface temperature (SST) and continental precipitation using the extended ERA5 data set from 1950-2019 \citep{hersbach_era5_2019,bell_era5_2020} provided by the European Centre for Medium-Range Weather Forecasts (ECMWF) as a state-of-the-art replacement of the ERA‐Interim reanalysis \citep{dee_era-interim_2011}. In general, trends and low frequency variability of surface temperature and humidity are represented well for the period 1979-2019 \citep{hersbach_era5_2019, simmons_low_2021}. However, \cite{simmons_low_2021} noticed strong biases in temperature and humidity records for Central and East Africa in the 1950s. Via inspection, we found that East African precipitation records in particular are strongly biased during that period towards much higher values. In order to avoid providing our Theta model extension with an incorrect starting point of the time series, we remove the first 10 years, providing us data from 1960 to 2019. Furthermore, the SVD of the covariance matrix is a rather memory intensive numerical operation, which is why we limit the domain of interest from \ang{40;;}~S to \ang{60;;}~N with a \ang{1;;}x\ang{1;;} spatial resolution, guaranteeing most of the continents to be included into the analysis. In total, the fields of SST and continental precipitation cover $n_A=\num{21816}$ and $n_B =\num{14184}$ grid points, respectively. To align the different temporal scales of highly variable precipitation and slow-varying SST and to filter out the high frequency signals, we smooth both data sets with a 6-month moving average, for each month taking into account the 3 preceding and the 2 following months. We choose this particular time window to effectively filter out the semi-annual cycles present in the geophysical observations. We then normalise both data fields by dividing each grid point by its temporal standard deviation to ensure equal spatial importance. In a non-standardised MCA analysis, regions of high rainfall, like the tropics, would dominate the covariance patterns, since annual rainfall in the mid-latitudes and subtropics is typically much lower. After normalisation, we weight the data points located on the regular \ang{1;;}x\ang{1;;} grid according to their associated area on a sphere \citep{north_sampling_1982} by multiplying each grid point with $\sqrt{\cos(\varphi_j)}$, $\varphi_j$ being the latitude at grid point $j$.

Climate indices of monthly means, used in this study for the sake of comparison,
are downloaded from the websites of the National Oceanic and Atmospheric Administration (NOAA; \url{https://psl.noaa.gov/data/climateindices/list/}), namely the Oceanic Niño Index (ONI) provided by the NOAA Climate Prediction Center, the Pacific Decadal Oscillation \cite[PDO,][]{mantua_pacific_2002} and the Atlantic Meridional Mode \cite[AMM,][]{chiang_analogous_2004}. We calculate the ENSO Modoki Index \cite[EMI,][]{ashok_nino_2007} according to $\text{EMI} = R_A - \sfrac{1}{2} R_B - \sfrac{1}{2} R_C$ where $R$ denotes the area-averaged ERA5 SST anomalies over the regions $A$ (\ang{165;;}E - \ang{140;;}W, \ang{10;;}S - \ang{10;;}N), $B$ (\ang{110;;}W - \ang{70;;}W, \ang{15;;}S - \ang{5;;}N) and $C$ (\ang{125;;}E - \ang{145;;}E, \ang{10;;}S - \ang{20;;}N), respectively. Furthermore, we define an oceanic warming index (OWI) as the area average of ERA5 SST anomalies (\ang{40;;}~S - \ang{60;;}~N) using the same latitude area weighting as for the MCA data preparation. Finally, all indices are smoothed using a 6-month moving average in alignment with our data pre-processing. We further center and max-normalise all indices for better comparison with our obtained PCs.

\section{Results}\label{sec:results}
In the following, we discuss the results from the synthetic experiments before investigating the results of the analysis of SST and continental precipitation.

\subsection{Synthetic Data}\label{sec:results_synthetic}
Before we start the investigation of the synthetic data, we apply standard and complex MCA as a baseline study on Gaussian white noise. By comparing the individual singular values as well as their sum obtained for each method (Tab.~\ref{tab:singular_values}), we observe that complex MCA consistently yields higher covariance both for each mode individually and overall, as a result of the additional time-lagged cross-covariance between field $A$ and $B$. Therefore, when comparing standard and complex MCA in the following, we will make use of the singular values directly instead of the covariance fraction $\gamma_k$ in order to assess the explained covariance by each mode.

\subsubsection{Experiment I: Lagged signals}
\begin{figure*}
    \noindent\includegraphics[width=\textwidth, angle=0]{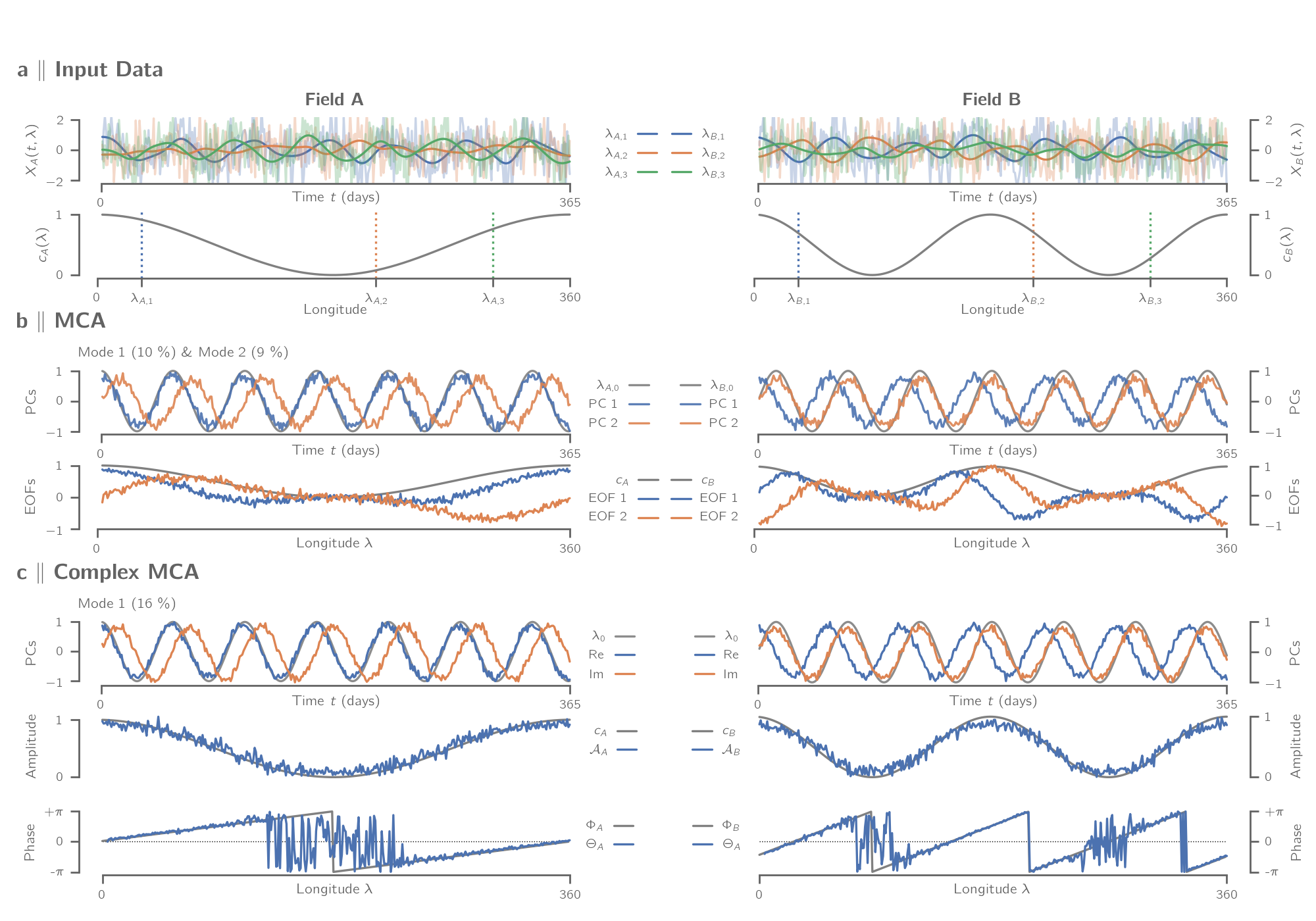}
    \caption{Results of Experiment I. \textbf{(a)} Illustration of synthetic data fields $A$ (left) and $B$ (right) showing \textbf{(top)} the temporal evolution $X_s(t,\lambda)$ for some selected longitudes $\lambda$ smoothed via a weighted running average using a Gaussian kernel with a standard deviation of $7$ days and \textbf{(bottom)} the associated modulation patterns $c_s(\lambda)$ for both fields, respectively. \textbf{(b)} Standard MCA solution showing mode 1 (explaining $10\%$ of the total covariance) and mode 2 ($9\%$) with \textbf{(top)} the associated PCs and \textbf{(bottom)}. \textbf{(c)} Complex MCA solution showing mode~1 ($16\%$) with \textbf{(top)} the associated real (Re) and imaginary (Im) parts of the first PC, \textbf{(centre)} the spatial amplitude functions and \textbf{(bottom)} the phase functions. For comparison the denoised signal $X_s(t, \lambda_{s,0}=0)$ (grey line, $\lambda_0$), the spatial modulation pattern $c_s(\lambda)$ (grey line, $c_s$) and the theoretical phase shift $\Phi_s = k_s \lambda - \phi_s$ (grey line, $\Phi_s$) are shown for both fields, respectively. The phase functions indicates the phase shift that has to be applied to the time series at a specific longitude $\lambda$ in order to obtain the real part of the non-shifted PC (shown in top panel). Percentages in parenthesis represent the covariance fraction $\gamma_k$. All EOFs and spatial amplitude functions are max-normalised for the sake of readability.}
    \label{fig:toydata1}
\end{figure*}

Applying standard MCA (denoted by subscript $std$), the travelling wave is split into two modes the first explaining $10\%$ of the shared covariance, and the second $9\%$ (Fig.~\ref{fig:toydata1}b). This is not surprising since the singular vectors are orthogonal and the associated PCs are uncorrelated, forcing the signals into two distinct modes following a sine and a cosine, which are perfectly uncorrelated for a full number of cycles and thus orthogonal. While the first two PCS correctly indicate the temporal evolution of the field following a sine/cosine, the EOFs fail to provide a clear and easy interpretation of the associated spatial structures. In particular, without further \textit{a priori} knowledge of the expected covarying structures, it seems a trying exercise to deduce the travelling wave from the obtained PCS and EOFs. Apart from that the remaining modes do not show any more distinct patterns, basically representing noise.

In comparison, complex MCA (denoted by subscript $com$) allows to represent the travelling wave by a single mode (Fig.~\ref{fig:toydata1}c). Additionally, the varying strength of the signal at different longitudes is clearly captured by the spatial amplitude for both fields. Another advantage in the interpretation compared to standard MCA is the spatial phase, which shows the migration of the signal to higher longitudes. Identical phase values at different longitudes indicate correlation while a phase shift of $\pi$ represents anti-correlation. We note that we have not used the Theta model extension here, since the signals in this experiment are stationary and spectral leakage due to non-integer number of periods can be assumed to be marginal. Furthermore, comparing the singular values $\sigma_{com,1} > \sigma_{std,1} + \sigma_{std,2}$ (Tab.~\ref{tab:singular_values}) indicates that the complex mode 1 accounts for more covariance and thus captures more of the travelling wave than the combined mode 1 and 2 of the standard solution. This makes sense, since the contributions of longitudes to the individual EOFs 1 and 2 of the standard solution will decrease for phase shifts which are not equal to $0$ (i.e. sine) or $\sfrac{\pi}{2}$ (i.e. cosine).
As a pitfall of complex MCA, however, it should be noted that longitudes whose spatial amplitude function is very low tend to have noisy phase function values. In our experiment, this is obvious for longitudes of both fields where $c_s(\lambda) \approx 0$ and the associated phase function does no longer follow the positive linear trend. At these longitudes, the max-normalised spatial amplitude function falls below $0.2$ for both fields, serving as a general orientation for the consideration of phase function values in the following analysis. In general, it is therefore advisable not to consider regions with low amplitudes.

\begin{figure*}
    \centering
    \includegraphics[width=\textwidth, angle=0]{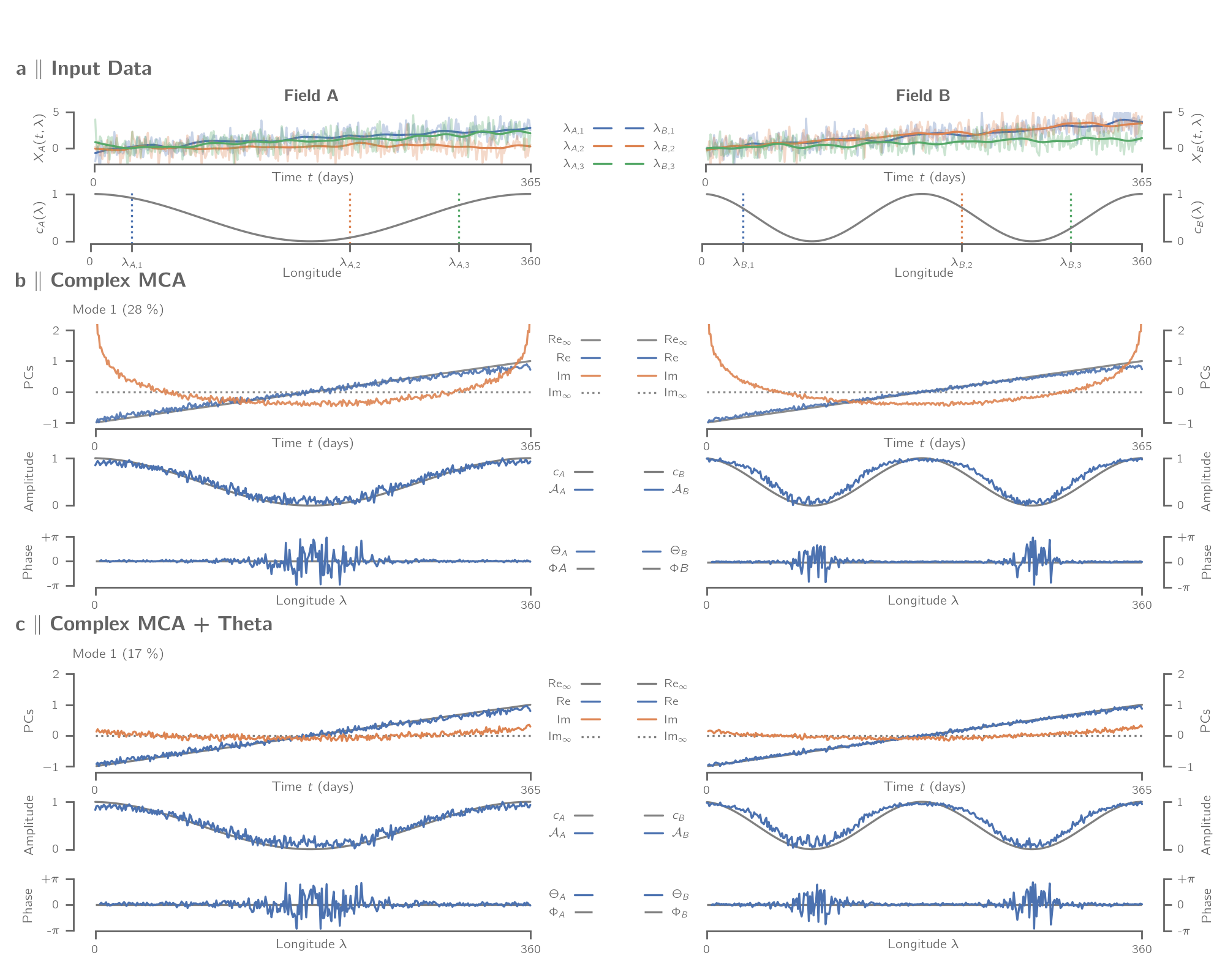}
    \caption{Results of Experiment II. \textbf{(a)} Illustration of synthetic data fields $A$ (left) and $B$ (right) showing \textbf{(top)} the temporal evolution $X_s(t,\lambda)$ for some selected longitudes $\lambda$ smoothed via a weighted running average using a Gaussian kernel with a standard deviation of $7$ days and \textbf{(bottom)} the associated modulation patterns $c_s(\lambda)$ for both fields, respectively. \textbf{(b)} Complex MCA solution showing mode 1 (explaining $28\%$ of the total covariance) with \textbf{(top)} the associated real and imaginary parts of the first PC, \textbf{(centre)} the spatial amplitude functions and \textbf{(bottom)} the phase functions. \textbf{(c)} As in \textbf{(b)} but showing mode 1 ($17\%$) for the Theta extended complex MCA. For comparison theoretical real and imaginary part of a infinite linear trend (grey lines, Re$_{\infty}$ and Im$_{\infty}$), the spatial modulation pattern $c_s(\lambda)$ (grey line, $c_s$) and the theoretical phase shift $\Phi_s = 0$ (grey line, $\Phi_s$) are shown for both fields, respectively. Percentages in parenthesis represent the covariance fraction $\gamma_k$. All spatial amplitude functions are max-normalised for the sake of readability.}
    \label{fig:toydata2}
\end{figure*}

\subsubsection{Experiment 2: Trends and non-cyclic signals}
As discussed in Sec.~\ref{sec:cmca}, non-periodic behaviour usually leads to undesired boundary effects caused by spectral leakage in the frequency domain. In our experiment, this effect is clearly evident for both PCs of the complex MCA (denoted by subscript $com$) (Fig.~\ref{fig:toydata2}b). Note that although the boundary effects seem to occur only in the imaginary parts of the PCs, this may not be the case in general. The Theta extended complex MCA (denoted by subscript $thc$), on the other hand, successfully mitigates the boundary effects of the PCs (Fig.~\ref{fig:toydata2}c), where we have set the Theta period $T_s=1$ as the time series have no seasonality. Nevertheless, the spatial patterns are similar for both methods and can hardly be distinguished visually. Considering only spatially static fields in this experiment, the phase function is constant for both fields, with exception at longitudes of low amplitude values as already mentioned in Experiment I.
Examining the singular values, we notice that $\sigma_{com,1} >> \sigma_{thc,1} \approx \sigma_{std,1}$ (Tab.~\ref{tab:singular_values}), indicating the increased covariance of the standard Hilbert transform due to boundary effects compared to the Theta model extended Hilbert transform. This implies that boundary effects created by the Hilbert transforms can strongly affect correlations and, depending on their magnitude, lead to a severe "inflation" of the singular values. As a consequence, the boundary effects can appear as parts of the first modes, thus completely misleading the interpretation of the results.

More generally, the frequency spectrum of a linear trend on a given interval is typically broadband and thus the analytical signal constructed by the Hilbert transform has not a physical interpretation in terms of characteristic frequencies. Therefore, the phase function cannot be interpreted in terms of a physical phase shift. The only exception is for a phase shift of $\theta=0; \pm \pi$ since these relative phase shifts represent correlating and anti-correlating signals disregarding the mathematical nature of the phase. A fundamental consequence of this is that for modes whose PC is broadband (e.g. a trend), only correlating ($\theta \approx 0$) as well as anti-correlating ($\theta \approx \pm \pi$) patterns should be considered.

\begin{table*}[t]
\caption{Singular values of MCA (Standard), complex MCA (Complex) and Theta extended complex MCA (Complex + Theta) obtained for purely white noise random fields (Noise) and the synthetic experiments described in Section~\ref{sec:results_synthetic} (Experiment I and II). Covariance fraction $\gamma_k$ for mode $k$ (in $\%$ shown within parenthesis) is calculated according to Eq.~\eqref{eq:cf}. Some exemplary modes (\textbf{bold}) are depicted in Fig.~\ref{fig:toydata1} and~\ref{fig:toydata2}. Total cumulated covariance is shown in the last row.}
\label{tab:singular_values}
\centering
\small
% For LaTeX tables use
\begin{tabular*}{\textwidth}{@{\extracolsep\fill}llllllll}
\hline\noalign{\smallskip}
\multicolumn{1}{c}{} &
  \multicolumn{2}{l}{Noise} &
  \multicolumn{2}{l}{Experiment I} &
  \multicolumn{3}{l}{Experiment II} \\
\hline\noalign{\smallskip}
Mode & Standard & Complex & Standard & Complex & Standard & Complex &  Complex + Theta \\
\noalign{\smallskip}\hline\noalign{\smallskip}
$\sigma_1$ & $2.58$ ($1.0$) & $8.37$ ($1.3$) & $\mathbf{27.43}$ ($9.8$) & $\mathbf{102.77}$ ($16.1$) & $89.1$ ($31.4$) & $\mathbf{179.82}$ ($28.2$) & $\mathbf{93.69}$ ($17.0$) \\
$\sigma_2$ & $2.53$ ($0.9$) & $8.11$ ($1.3$) & $\mathbf{24.05}$ ($8.6$) & $6.97$ ($1.1$) & $1.95$  ($0.7$) & $6.19$ ($1.0$) & $6.18$ ($1.1$) \\
$\sigma_3$ & $2.50$ ($0.9$) & $7.84$ ($1.2$)  & $2.21$ ($0.8$) & $6.82$ ($1.1$) & $1.92$ ($0.7$) & $5.96$ ($0.9$) & $5.97$ ($1.1$) \\
$\vdots$ & $\vdots$ & $\vdots$ & $\vdots$ & $\vdots$ & $\vdots$ & $\vdots$ \\
\noalign{\smallskip}\hline
$\sum_j \sigma_j$ & $269.91$ & $629.95$  & $279.96$ & $640.02$ & $284.1$ & $637.97$  & $551.16$ \\
\noalign{\smallskip}\hline
\end{tabular*}
\end{table*}

\subsection{Climate Data: SST \& Continental Precipitation}\label{sec:results_climate}
\begin{figure*}
    \noindent\includegraphics[width=\textwidth]{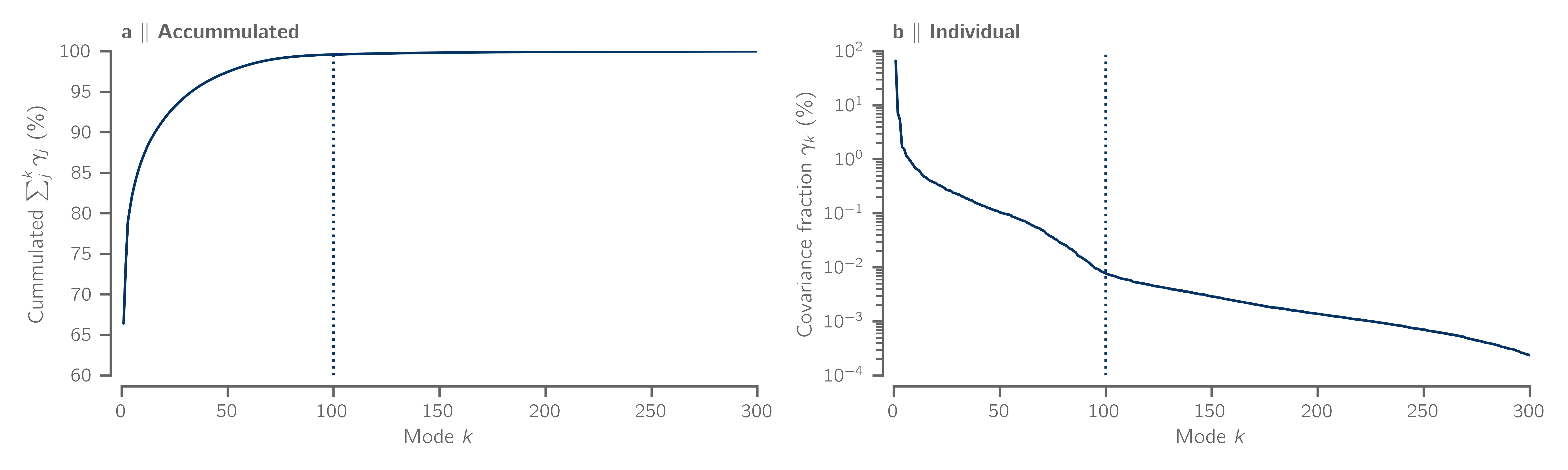}
    \caption{Described covariance of the first 300 modes of complex MCA applied to SST and continental precipitation showing the (a) accumulated and (b) individual share. Dashed line estimates the boundary of suspected "noisy" modes.}
    \label{fig:mca_described_variance}
\end{figure*}

\begin{figure*}
    \noindent\includegraphics[width=\textwidth]{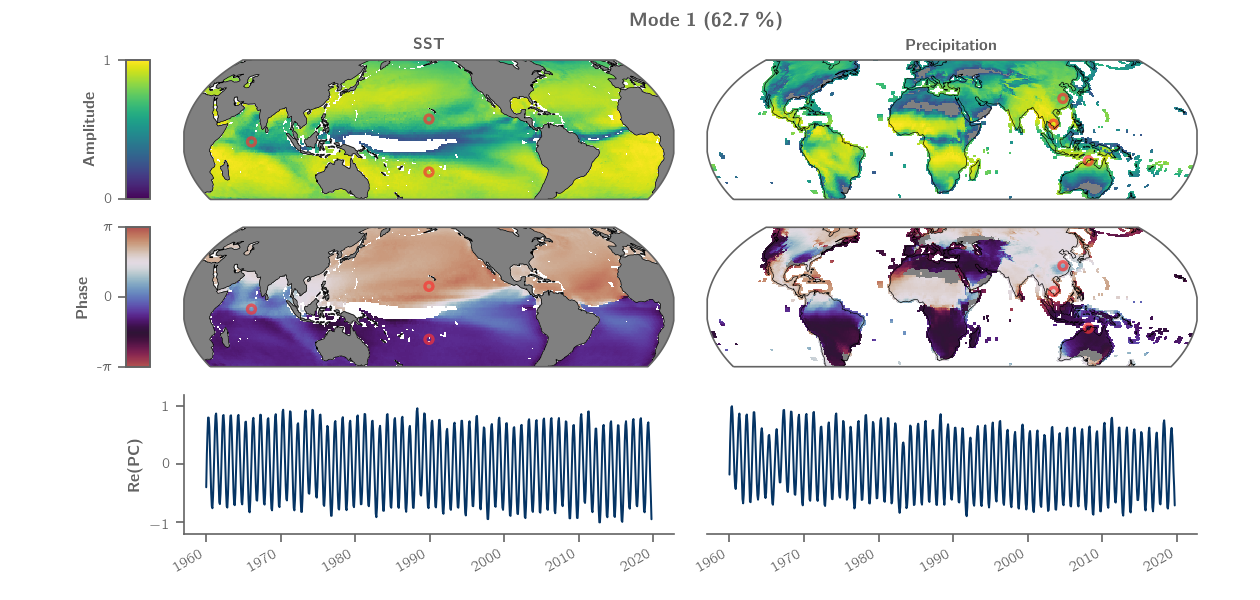}
    \caption{Complex Varimax-rotated MCA of \textbf{(left)} SST and \textbf{(right)} continental precipitation showing mode 1 and its relative importance indicated by the covariance fraction. \textbf{(top)} The amplitude functions show the regions predominantly contributing to the mode. \textbf{(centre)} The phase functions depict the relative phase shifts with respect to \textbf{(bottom)} the corresponding real part of the PC, where $0$ (blue) means correlation and $\pm \pi$ (red) shows anti-correlation. For each grid point, the corresponding PC can be computed from the given phase value by applying the negative phase shift to the PC of phase $0$ (bottom).  The amplitude functions, PCs and indices shown are all max-normalised for the sake of comparability. In both amplitude and phase function regions with a max-normalised amplitude below $0.25$ are masked out. Red circles mark North Pacific (\ang{170;;}~E, \ang{15;;}~N), Indian Ocean (\ang{70;;}~W, \ang{0;;}~N) and South Pacific (\ang{160;;}~E, \ang{20;;}~S) for SST and Poyang lake, China (\ang{116.3;;}~W, \ang{29.1;;}~N), Phnom Penh, Cambodia (\ang{104.9;;}~W, \ang{11.6;;}~N) and Darwin, Australia (\ang{130.8;;}~W, \ang{12.5;;}~S) for precipitation. The phase shifted PCs of these locations are examined in Fig.~\ref{fig:seasonal_shifts}.}
    \label{fig:results_modes1}
\end{figure*}

We apply Theta extended, complex MCA to SST and continental precipitation using a Theta period $T_s=12$ to account for the seasonal cycle. The dimensionality of the problem can be greatly reduced with the first $72$ modes explaining more than $99 \%$ of the lagged covariance (Fig.~\ref{fig:mca_described_variance}a). In order to simplify the obtained patterns and to increase the physical meaning of the results (Sec.~\ref{sec:rmca}), we rotate the first \num{150} modes explaining \SI{99.82}{\percent} of the lagged covariance. Our decision to rotate \num{150} modes is motivated by the idea of retaining as much information as possible without including the noise which dominates the higher modes. Since we observe a marked drop in singular values at about the mode number \num{100} followed by an exponential decrease in singular values (Fig.~\ref{fig:mca_described_variance}b), we guess that there is a negligible information content in higher modes. Moreover, the quality of the reconstructed signal is rather independent of the exact number of rotated modes, with $150\pm 50$ yielding basically identical results for the first 6 modes for both Varimax orthogonal and Promax $p=2$ oblique rotation. Therefore, we will restrict our discussion in the following to the first 6 modes. In order to address the question of the rotation method to be chosen, we note that Promax oblique rotation performs better when simple structures are present and correlations among PCs are high \citep{finch_comparison_2006}. We expect our first mode to be dominated by the shared dynamics of the seasonal cycle of both SST and continental precipitation, which is indeed what we find for both Promax (not shown here) and Varimax solution (Fig.~\ref{fig:results_modes1}). This mode, however, is fairly global and as such does not represent a simple structure. Furthermore, we observe that the correlations among the first six Promax-rotated PCs range from -0.15 to 0.13 only, underlining that the Promax oblique solution is close to orthogonal and differences to the Varimax solutions only marginal, at least for the first six modes. Since the Promax oblique solution seemingly does not provide a better results, we opt for the somewhat simpler Varimax orthogonal rotation. In the following, we will discuss the Varimax-rotated modes by investigating the real part of the PCs, the spatial amplitude and phase functions for both fields, SST and continental precipitation, respectively.

In our representations of the modes, we remove non-significant, "noisy" phase values (see Sec.~\ref{sec:results_synthetic}) by masking out regions in the spatial amplitude and phase function exhibiting a max-normalised amplitude of $<0.25$.

\begin{figure*}
\noindent\includegraphics[width=\textwidth]{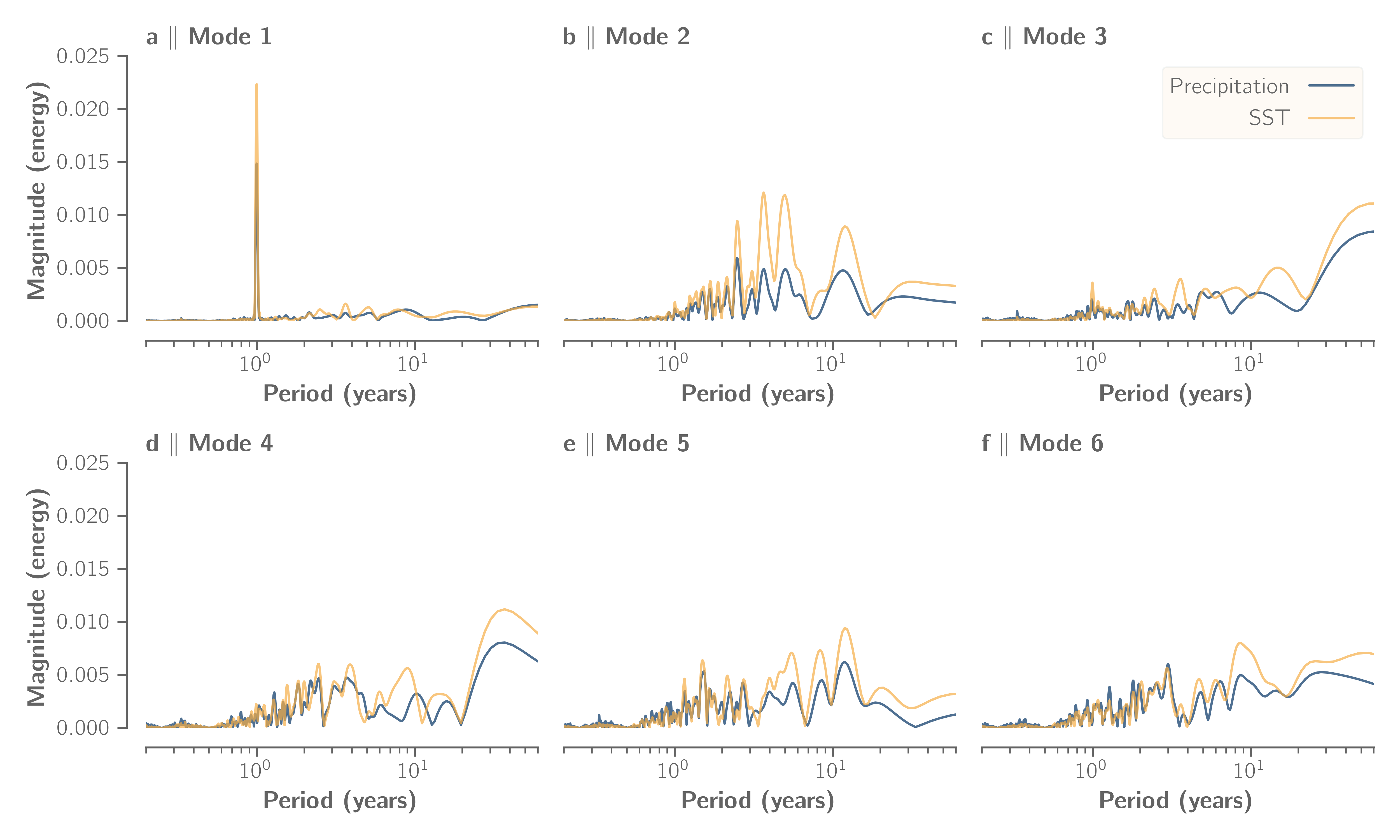}
\caption{Magnitude spectrum of the first 6 PCs considering only the real part and using a Hanning window.}
\label{fig:mca_fft}
\end{figure*}

Mode 1 describes \SI{62.7}{\percent} of the covariance between SST and continental precipitation clearly showing the annual cycle (Figs.~\ref{fig:results_modes1} and \ref{fig:mca_fft}a). As expected, the annual cycle shows itself in both variables on a global scale, with the exception of the equatorial ocean, where the seasonal SST variations are only weak. The phase function correctly identifies the anti-correlation between the northern and southern oceans. It also suggests that the eastern equatorial Pacific and the equatorial Indian Ocean nevertheless show a weak seasonal signal, which is, however, positively phase shifted relative to the rest of the southern ocean. On the continents, precipitation dominates mainly in monsoon areas and the tropics. The phase function illustrates the division into June to August (white) and December to February (black) dominated rainfall systems and identifies corresponding transition zones, as e.g. over the South American rain forest and central North America. It also highlights some interesting dynamical regions which stand out of their respective environment, namely, the West and East coast of North America, the Mediterranean region, and to a lesser degree the East Asian monsoon region in China.

\begin{figure*}
    \centering
    \includegraphics[width=\textwidth]{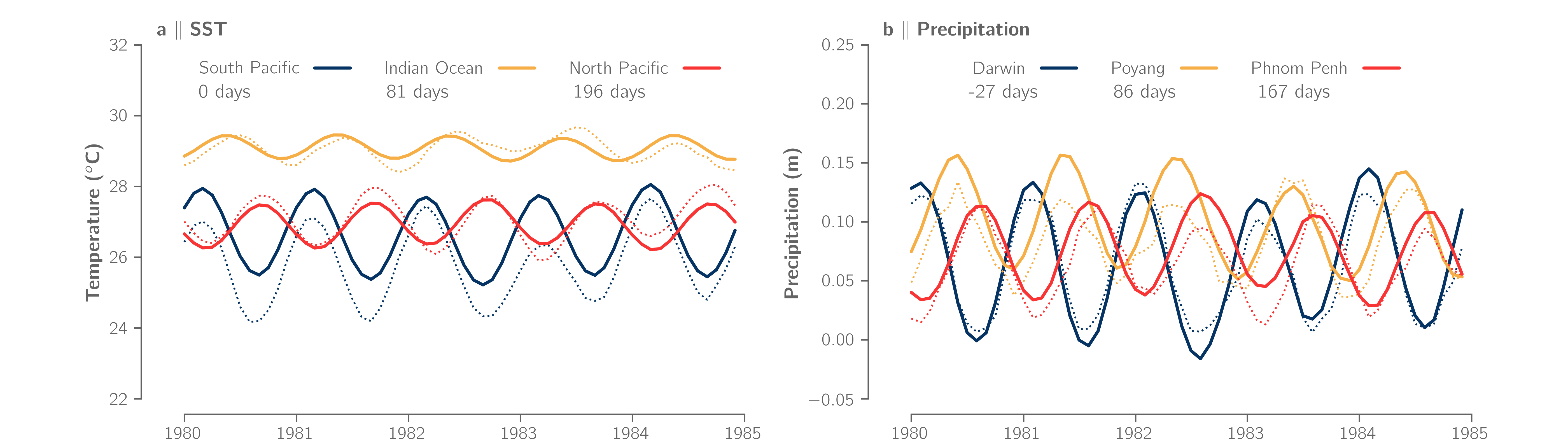}
    \caption{Comparison between 6-month moving average of ERA5 SST and continental precipitation (dotted lines) and the reconstructed time series based on mode 1 only (thick lines) for the exemplary locations defined in Fig.~\ref{fig:results_modes1} covering the years 1980 to 1985. Days in legend refer to the time shift $\tau$ of the individual locations with respect to the SST of the South Pacific derived from the spatial phase functions (Fig.~\ref{fig:results_modes1}).}
    \label{fig:seasonal_shifts}
\end{figure*}

The obtained mode provides an instructive example to highlight the benefits of the complexified approach due to the mode's corresponding narrow-band frequency spectrum  (Fig.~\ref{fig:mca_fft}a). Using the dominant periodicity of mode~1, $T_1 = 12$ months, the interpretation of a phase shift $\theta$, given by the spatial phase function, as time lag $\tau$ is straightforward and can be computed via $\tau = \sfrac{T_1\theta}{2\pi}$. Doing this for some exemplary locations (denoted by red circles in Fig.~\ref{fig:results_modes1}), we observe that the seasonal SST maximum of the North Pacific follows the one in the South Pacific by $196$ days, that is, being almost perfectly anti-correlated (Fig.~\ref{fig:seasonal_shifts}). The relative time shifts of the SST and precipitation maximum of $81$ days in the Indian Ocean and $86$ days at Poyang lake, China, is close to $3$ months and as such translates to a phase shift of about $-\sfrac{\pi}{2}$, something that cannot be picked up as feature within one mode when using standard MCA. It should also be noted that although the sampling frequency of SST and precipitation is monthly, the phase function is continuous and thus allows to infer time lags on shorter time scales. This is, for instance, the case for the seasonal precipitation maximum in Darwin, Australia, which precedes the seasonal SST maximum of the South Pacific by $27$ days.

\begin{figure*}
    \noindent\includegraphics[width=\textwidth]{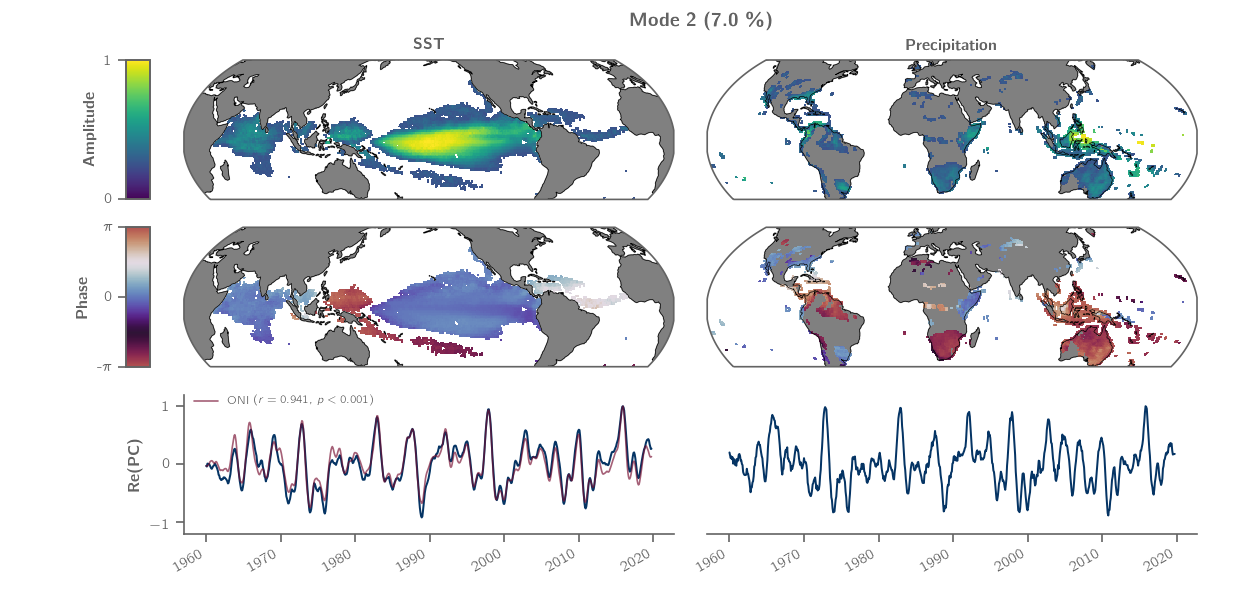}
    \caption{As in Fig.~\ref{fig:results_modes1} but showing mode 2 corresponding to the Oceanic Niño Index (ONI) provided by NOAA Climate Prediction Center as described in Section~\ref{sec:data_climate}. The Spearman correlation between ONI and the real part of SST PC 2 is a $r=0.941$ with $p\text{-value }<0.001$.}
    \label{fig:results_modes2}
\end{figure*}

\begin{figure*}
    \noindent\includegraphics[width=\textwidth]{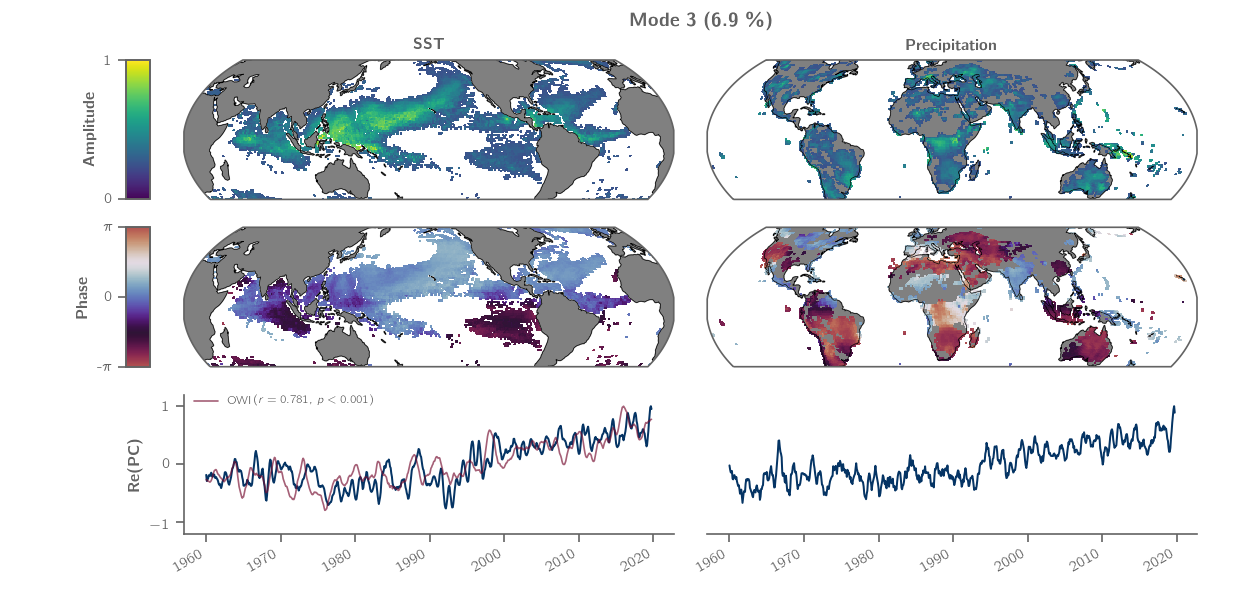}
    \caption{As in Fig.~\ref{fig:results_modes1} but showing mode 3 corresponding to the oceanic warming index (OWI) as described in Section~\ref{sec:data_climate}. The Spearman correlation between the OWI and the real part of SST PC 3 is $r=0.781$ with a $p\text{-value }<0.001$.}
    \label{fig:results_modes3}
\end{figure*}

\begin{figure*}
    \noindent\includegraphics[width=\textwidth]{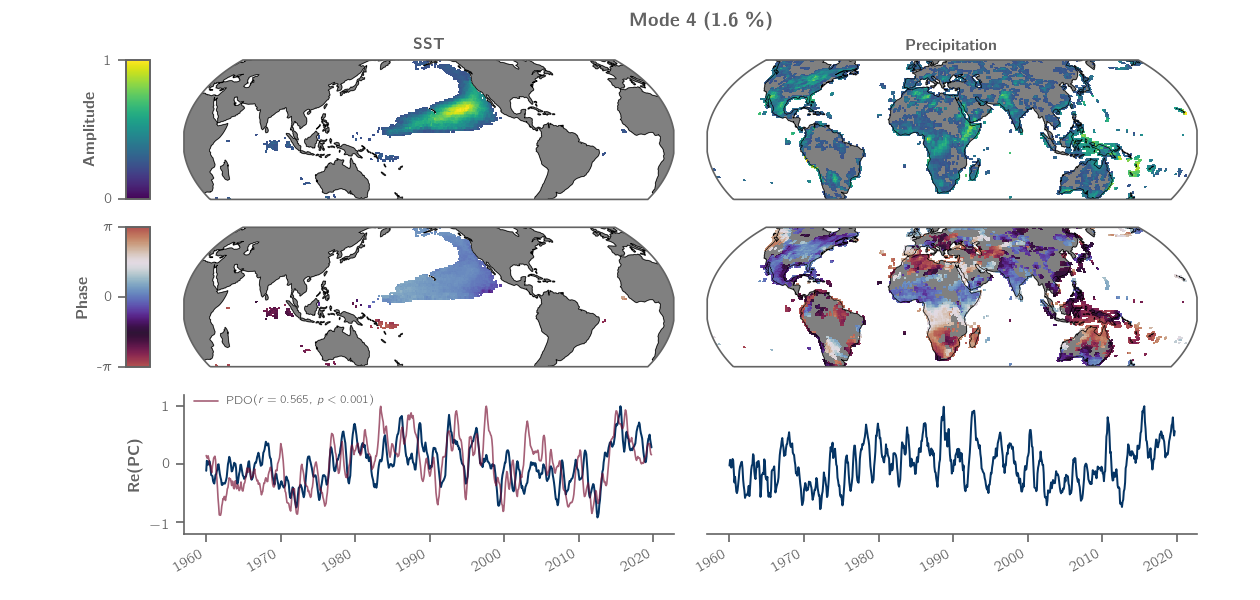}
    \caption{As in Fig.~\ref{fig:results_modes1} but showing mode 4 corresponding to the Pacific Decadal Oscillation \cite[PDO,][]{mantua_pacific_2002} as described in Section~\ref{sec:data_climate}. The Spearman correlation between the PDO and the real part of SST PC 4 is $r=0.565$ with a $p\text{-value }<0.001$.}
    \label{fig:results_modes4}
\end{figure*}

Mode 2 (\SI{7.0}{\percent}) can be clearly associated to the ocean-atmosphere phenomenon of El Niño - Southern Oscillation (ENSO)
(Fig.~\ref{fig:results_modes2} and \ref{fig:mca_fft}b). During \textit{El Niño}, higher SST in the central and eastern Pacific and lower SST in the western Pacific positively correlate with heavier rainfall within a narrow band along the west coast and the southeastern coast of South America \citep{tedeschi_influences_2013},
the east and west coasts of North America \citep{ropelewski_north_1986}, the East Asian monsoon region \citep{wen_direct_2019} and the Horn of Africa \citep{indeje_enso_2000}. At the same time, precipitation decreases in northern South America \citep{tedeschi_influences_2013}, Oceania \citep{dai_global_2000}, South Africa \citep{gaughan_inter-_2016} and in the Indian monsoon region \citep{cherchi_influence_2013}. During \textit{La Niña} (the counterphase of \textit{El Niño}), these correlations are reversed. Recently, similar teleconnections have been identified via event coincidence analysis \citep{wiedermann_differential_2021}. For a current summary of established ENSO-related rainfall patterns during El Niño and La Niña see \citet{lenssen_seasonal_2020}.

Our result also shows the dynamical link between ENSO in the Pacific Ocean and the Indian Ocean \citep{krishnamurthy_variability_2003}, the South China Sea \citep{klein_remote_1999} and the Tropical North Atlantic \citep{enfield_tropical_1997,saravanan_interaction_2000,alexander_influence_2002,chiang_tropical_2002} in accordance with previous studies. Interestingly, it was found that the ENSO related SST teleconnections in the remote oceans often occur with some delays, with the Indian Ocean typically peaking $\sim3$ months and the South China Sea and tropical North Atlantic $\sim5$ months after the SST peak in the Pacific ENSO region \citep{enfield_tropical_1997,klein_remote_1999,saravanan_interaction_2000}. The mechanism behind these lagged responses, known as the atmospheric bridge, is based on the characteristic atmospheric circulation during El Niño which causes changes in cloud cover and evaporation over the remote oceans, leading to increased net heat flux and SSTs \citep{lau_role_1996,klein_remote_1999}. However, due to the broadband frequency spectrum of the SST PC (Fig.~\ref{fig:mca_fft}b), with most of the energy contained at four different peaks around $2.5$, $3.5$, $5$ and $11$ years, the phase cannot simply be translated into a time shift. Nevertheless, the tropical North Atlantic clearly exhibits more positive phase values compared to the Pacific El Niño region, therefore indicating to be positively time shifted relative to the Pacific.

Mode 3 (\SI{6.9}{\percent}) represents global warming and the associated changes in precipitation patterns (Fig.~\ref{fig:results_modes3}). The warming SST patterns clearly emerge in all major ocean basins, although more pronounced in the northern hemisphere due to the asymmetric response of the northern and southern trade winds to global warming \citep{xie_global_2010}. We also note a pronounced warming of the western part of both, the Pacific and the Atlantic basins, both regions of enhanced ocean heat transport \citep{xie_global_2010,hannachi_archetypal_2017}. Similar to these oceanic trends, we also observe global trends in the precipitation patterns, with decreasing rainfall over the Mediterranean, South Africa, Australia, South America and parts of western North America. There seems to be also a decrease of rainfall over the west Asian monsoon region. In contrast to that, the results suggest increased precipitation over the Indian monsoon region as well as some localised regions in Europe, South and North America. These results are largely in agreement with studies based on observational data \citep{gu_spatial_2015} and, more recently, on CMIP5 climate simulations \citep{giorgi_response_2019}. Finally, it should be stressed, that both PCs, SST and precipitation, do only provide a meaningful interpretation for phases $\theta \approx 0;\pm\pi$ (correlation, anti-correlation), due to their broadband frequency spectra (Fig.~\ref{fig:mca_fft}c). For phase shifts different from that, no clear conclusions can be drawn, as it is the case e.g. for equatorial Africa where the phase shift is approximately $+\pi/2$.

Mode 4 (\SI{1.6}{\percent}) shows a slow-oscillating pattern of SST in the northeastern Pacific correlating with localized precipitation patterns distributed over all continents (Fig.~\ref{fig:results_modes4}). The typical spatial SST pattern is known as the Pacific Decadal Oscillation (PDO) \citep{mantua_pacific_1997} and a well-established climate index. A combination of various processes originating in the tropics and extra-tropics has been proposed as the physical source of the PDO \citep{newman_pacific_2016}, with ENSO and PDO likely responding to the same forcing function \citep{pierce_role_2002}. Our analysis, however, provides a means to disentangle ENSO and PDO related precipitation patterns, which are often similar for western North America \citep{hu_interferential_2009}, though reveal important differences e.g. for Australia, the Indian subcontinent or the African Sahel region. Yet, care must be taken when interpreting regions which have a phase shift different from $0,\pm \pi$. Although the PDO exhibits a strong spectral energy at about $~\SI{35}{yr}$, the mode contains also important features at about $\SI{1}{yr}$ to $\SI{4}{yr}$ (Fig.~\ref{fig:mca_fft}d) making the phase interpretation physically less clear.

\begin{figure*}
    \noindent\includegraphics[width=\textwidth]{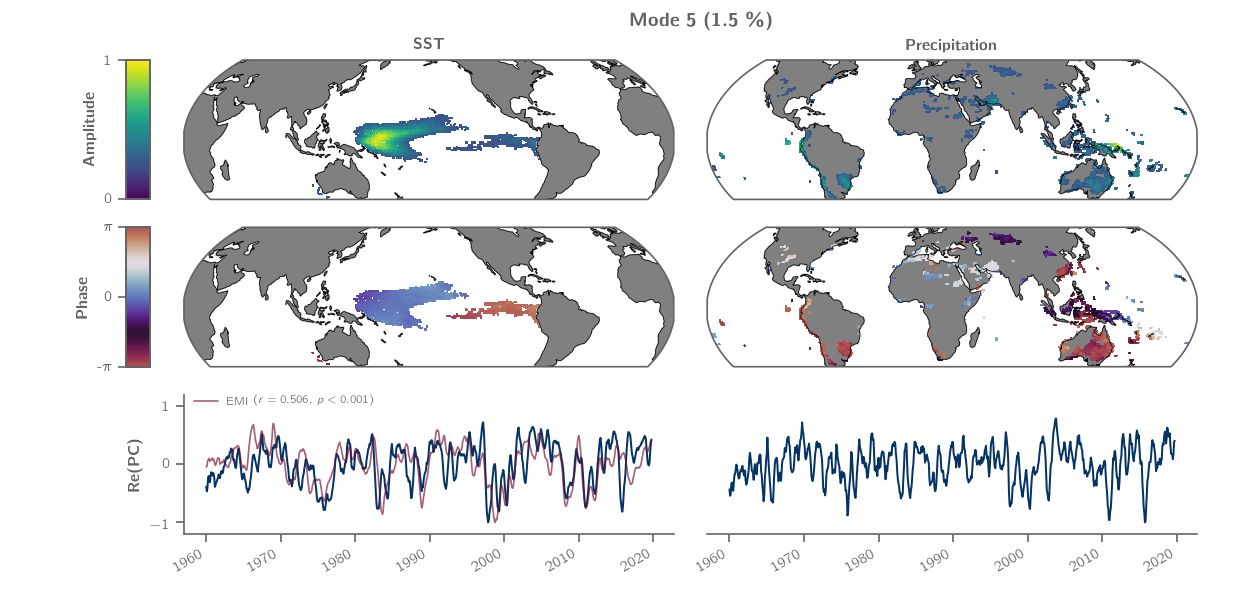}
    \caption{As in Fig.~\ref{fig:results_modes1} but showing mode 5 corresponding to the ENSO Modoki Index \cite[EMI,][]{ashok_nino_2007} as described in Section~\ref{sec:data_climate}. The Spearman correlation between the EMI and the real part of SST PC 5 is $r=0.506$ with a $p\text{-value }<0.001$.}
    \label{fig:results_modes5}
\end{figure*}

Mode 5 (\SI{1.6}{\percent}) describes an oscillating SST anomaly mainly limited to the central Pacific (Fig.~\ref{fig:results_modes5} and \ref{fig:mca_fft}e) describing El Niño Modoki \citep{ashok_nino_2007} and represented by the El Niño Modoki Index (EMI). Higher SST in the central Pacific and lower SST in the east Pacific correlate with reduced precipitation in the East Asian monsoon region \citep{feng_different_2011}, Australia \citep{taschetto_nino_2009}, parts of South America \citep{tedeschi_influences_2013} and South Africa \citep{ratnam_remote_2014} and vice versa. Some of these teleconnections have recently been revealed by event coincidence analysis \citep{wiedermann_differential_2021} though important links to e.g. the East Asian monsoon region were missing. Although our result suggests that continental rainfall in certain regions of Africa, Arabia and the Americas are phase-shifted El Niño Modoki expressions, future work has to show if these weak amplitudes are significant.

\begin{figure*}
\noindent\includegraphics[width=\textwidth]{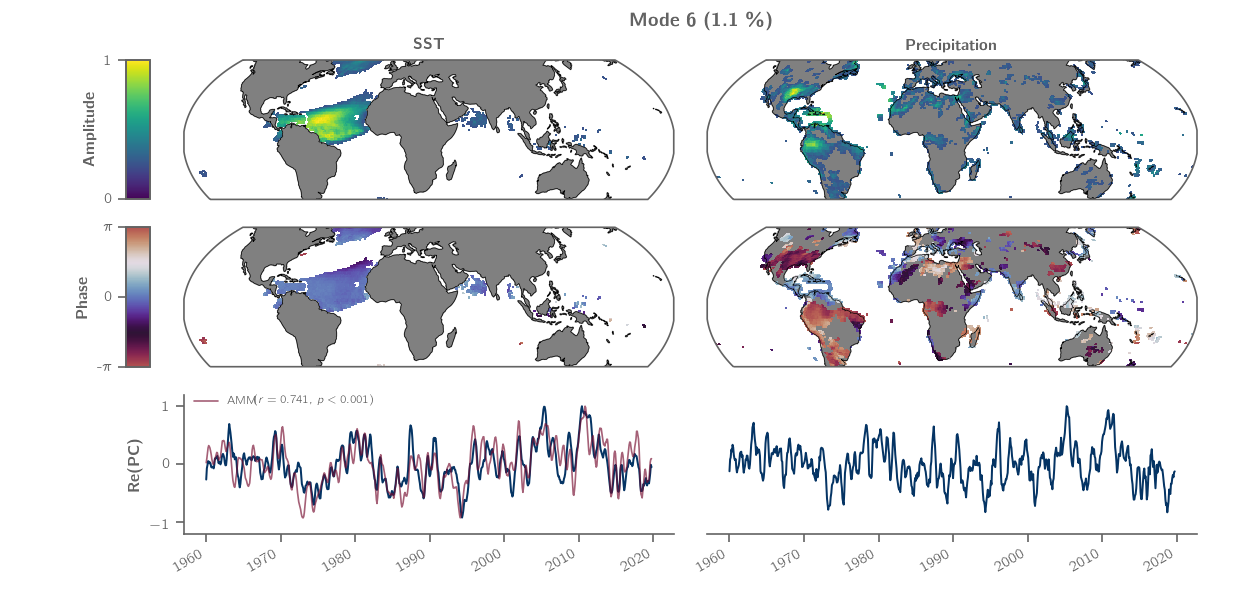}
\caption{As in Fig.~\ref{fig:results_modes1} but showing mode 6 corresponding to the Atlantic Meridional Mode \cite[AMM,][]{chiang_analogous_2004} as described in Section~\ref{sec:data_climate}. The Spearman correlation between the AMM and the real part of SST PC 6 is $r=0.741$ with a $p\text{-value }<0.001$.}
\label{fig:results_modes6}
\end{figure*}

Mode 6 (\SI{1.1}{\percent}) is characterised by a SST pattern concentrated to the tropical and subarctic North Atlantic (Fig.~\ref{fig:results_modes6} and \ref{fig:mca_fft}f). The SST pattern, representing the Atlantic Meridional Mode (AMM), is the dominant coupled ocean-atmospheric phenomenon in the tropical Atlantic \citep{servain_relationship_1999} and its impact on precipitation of the African Sahel zone, Central America and the northern South American continent are well known \citep{lamb_interannual_1986,martin_multidecadal_2014}. Only very recently, \citet{vittal_early_2020} found a link between the AMM and Indian summer monsoon rainfall, which they used to improve precipitation forecast. In addition, our result suggests more tele-connections between the AMM and precipitation variability over the Mediterranean, East Africa, the Congo basin, North America and the East Asian monsoon region, providing much potential of advancing local rainfall predictions in those areas. Due to the missing intrinsic time scale with no clear periodicity (Fig.~\ref{fig:mca_fft}f), only correlating and anti-correlating patterns should be interpreted. However, most of the regions identified by this mode satisfy being correlated or anti-correlated.

\section{Conclusion}\label{sec:conclusion}
Understanding the intricate climate system is a challenging task, that requires advanced statistical methods. Finding correlations among a set of different climate variables is complicated by the frequently present lagged responses of different variables to the same forcing. We show that complex rotated MCA provides a practical tool to single out such modes from a high-dimensional data space, even when the narrow-band assumption of the input signals is only partially satisfied.

By taking into account the spatial amplitude and phase function of the obtained complex modes, we obtain a simple approach to examine otherwise complicated spatial and temporal structures. Our synthetic experiments highlight that, in the case of phase-shifted signals, complex MCA can capture a more comprehensive and complete picture of the correlations present. However, they also show the sensitivity of the Hilbert transform to spectral leakage caused by boundary conditions of the given time series, e.g. when the time series clearly consists of a non-integer number of cycles and/or in the presence of trends. Since the stationarity assumption does often not hold (due to climate change), time series should generally be pre-processed when applying complex MCA.

Extending the time series via the optimised Theta model mitigates the effect of spectral leakage and produces physically reasonable PCs. This procedure allows us to resolve trends and non-cyclic signals, although for trends the obtained spatial phase function has a simple physical interpretation only for correlating and anti-correlating patterns. Nevertheless, this approach provides a means of applying complex MCA without the need to detrend the time series of interest. Moreover, excluding the first mode, the original fields can be reconstructed without the seasonal cycle, providing an advanced tool to pre-process time series containing non-stationary and non-linear seasonal features.

A general caveat in complex MCA is the fact that the phase function loses its interpretation for PCs with a broadband frequency spectrum. But although the spatial phase function has not always a simple physical interpretation for most of the modes due to their broadband frequency spectrum, complex MCA nevertheless can always be interpreted for the correlating and anti-correlating patterns.

Applying complex rotated MCA to SST and continental precipitation, we clearly identify the main shared dynamics in both variables, namely (i) the seasonal cycle, (ii) the canonical ENSO, (iii) the trends associated to global warming (iv) the PDO, (v) ENSO Modoki and (vi) the AMM. We also retrieve phase shifted signals between the two climate variables. While for the seasonal cycle these phase shifts can directly be translated into a time shift, the remaining modes generally do not lend itself to such a simple interpretation due to their broadband frequency spectra. But even without a precise equivalent as time lag, the phase function provides a means to identify regions of lagged correlations, for instance, between the SST of the Pacific and the tropical North Atlantic during El Niño events.  In addition, by focusing on narrow frequency bands only, one may detect frequency ranges over which the phase is not random, thus potentially uncovering more of dynamic structure of the individual modes.

Many of the obtained correlation patterns in SST and continental precipitation had already been evidenced by a multiplicity of different, partly regional studies. The great advantage of complex rotated MCA is that it allows to obtain all those patterns by a single analysis of the correlation of two geophysical variables at global scale in a more compact and easy-to-interpret way. Besides, our results also point out to new ocean-atmospheric teleconnections, that, to our knowledge, have not been reported, most notably for the PDO and the AMM.

Regarding future applications of complex rotated MCA, this method has the potential for shedding light in the investigation of seasonal and sub-seasonal phenomena, as well as for spatially propagating patterns. As future work, we plan to analyse the Madden-Julian oscillation. Additionally, complex rotated MCA could be used to evidence other connections between less studied variables, as for instance sea surface salinity, sea surface height, soil moisture, winds, etc, what has the potential of evidence new phenomena and novel aspects of existing or new teleconnections.

%%%%%%%%%%%%%%%%%%%%%%%%%%%%%%%%%%%%%%%%%%%%%%%%%%%%%%%%%%%%%%%%%%%%%
% ACKNOWLEDGMENTS
%%%%%%%%%%%%%%%%%%%%%%%%%%%%%%%%%%%%%%%%%%%%%%%%%%%%%%%%%%%%%%%%%%%%%
\section*{Acknowledgments}
This work is part of the Climate Advanced Forecasting of sub‐seasonal Extremes (CAFE) project and has been prepared in the framework of the doctorate in Physics of the Autonomous University of Barcelona. The authors gratefully acknowledge funding from the European Union's Horizon 2020 research and innovation programme under the Marie Skłodowska‐Curie grant agreement 813844, and also from the Spanish government through the ‘Severo Ochoa Centre of Excellence’ accreditation (CEX2019-000928-S). This work is a contribution to CSIC Thematic Interdiciplinary Platform TELEDETECT.

The authors acknowledge the Copernicus Climate Change Service and the Physical Sciences Laboratory of NOAA for providing the data. The authors would also like to thank J. Ballabrera-Poy for his valuable comments on a previous version of the paper.

%%%%%%%%%%%%%%%%%%%%%%%%%%%%%%%%%%%%%%%%%%%%%%%%%%%%%%%%%%%%%%%%%%%%%
% DATA AVAILABILITY STATEMENT
%%%%%%%%%%%%%%%%%%%%%%%%%%%%%%%%%%%%%%%%%%%%%%%%%%%%%%%%%%%%%%%%%%%%%
%
%
\section*{Data Availability Statement}
The ERA5 data used in this study are publicly available from the Copernicus Climate Data Store at \url{https://doi.org/10.24381/cds.f17050d7} as described in \citet{hersbach_era5_2019}. ERA5 preliminary extension from 1950-1978 is described in \citet{bell_era5_2020}. All climate indices used in this study are openly accessible. ENSO Modoki Index (EMI) is provided by the website of JAMSTEC at \url{http://www.jamstec.go.jp/virtualearth/general/en/}. The remaining indices can be found on the website of the Physical Sciences Laboratory of NOAA at \url{https://psl.noaa.gov/data/climateindices/list/}.

%%%%%%%%%%%%%%%%%%%%%%%%%%%%%%%%%%%%%%%%%%%%%%%%%%%%%%%%%%%%%%%%%%%%%
% APPENDIXES
%%%%%%%%%%%%%%%%%%%%%%%%%%%%%%%%%%%%%%%%%%%%%%%%%%%%%%%%%%%%%%%%%%%%%
%
\section*{Appendix}

\subsection{Definition of $\arctan\!2$}\label{app:atan2}
Every non-zero complex number in Cartesian coordinates, $z=x+iy$, can be transformed into polar complex coordinates, $z=ae^{i\theta}$, where $a=(x^2+y^2)^{1/2}$ is the amplitude, and $\theta$ the phase of $z$. For each $z\neq0$, the phase is only defined up to an integer multiple of $2\pi$, resulting in an infinite number of possible values. In order to construct a well-defined function $\theta(x,y)$, one typically limits the phase $\theta$ to $\left(-\pi,\pi \right]$. Then, the two-argument arctangent function $\arctan\!2(y,x)$ converts the values of $y$ and $x$ to the polar phase via:
\begin{equation*}
    \theta = \arctan\!2(y,x) =
    \begin{cases}
    \arctan(\frac{y}{x})        & \quad \text{if } x > 0 \\
    \arctan(\frac{y}{x}) + \pi  & \quad \text{if } x < 0 \text{ and } y \geq 0 \\
    \arctan(\frac{y}{x}) - \pi  & \quad \text{if } x < 0 \text{ and } y < 0 \\
    +\frac{\pi}{2}              & \quad \text{if } x = 0 \text{ and } y > 0 \\
    -\frac{\pi}{2}              & \quad \text{if } x = 0 \text{ and } y < 0 \\
    \text{undefined}            & \quad \text{if } x = 0 \text{ and } y = 0
  \end{cases}
\end{equation*}

\subsection{Finding the rotation matrix $\pmb{R}$}\label{sec:rotation_criterion}
Let us assume a complex loading matrix $\pmb{L}_r \in \mathbb{C}^{n \times r}$
containing only the first $r$ columns (modes) which are to be rotated and $n$ denoting the number of grid points. The number of grid points may be the sum of the grid points of two different fields $n = n_A + n_B$ as it is the case for MCA and described by Equation~\eqref{eq:loading_matrix} or simply the total number of grid points if only one field is considered as it is for PCA. In the following, we will drop the subscript $r$ in order to keep the notation simple and let $\widetilde{\cdot}$ denote the rotated solutions. Furthermore, $^*$ refers to the conjugate transpose of a matrix and $|\cdot|$ denotes the absolute value of a complex number.

\paragraph{Varimax orthogonal rotation}
The goal of Varimax rotation is to approximate simple structures \citep{thurstone_simple_1947} of the EOFs which is achieved by simplifying the columns of $\pmb{L}$ via an orthogonal rotation $\pmb{R}$. For this purpose, \citet{kaiser_varimax_1958} defines the simplicity $S_k$,
\begin{align}
    S_k = \frac{1}{n} \sum_{j=1}^n \left( |\widetilde{l}_{jk}|^2 \right)^2 - \frac{1}{n^2} \left( \sum_{j=1}^n |\widetilde{l}_{jk}|^2 \right)^2, \quad k=1,\dots,r
\end{align}
which measures the variance of the squared amplitude of the rotated loadings $\widetilde{l}_{jk}$. With increasing variance, the squared rotated amplitudes $|\widetilde{l}_{jk}|^2$ either become low or large, thus increasing simplicity.
The normalised simplicity $S$ then reads
\begin{align}
    S = \sum_{k=1}^r \left[ \frac{1}{n} \sum_{j=1}^n \left( \frac{|\widetilde{l}_{jk}|^2}{h_j^2} \right)^2 - \frac{1}{n^2} \left( \sum_{j=1}^n \frac{|\widetilde{l}_{jk}|^2}{h_j^2} \right)^2 \right],
\end{align}
where $h_j = \left( \sum_{k=1}^r |l_{jk}|^2 \right)^{1/2}$ represents the \textit{communality} of grid point $j$, which is the amount of variance of the $j$th gird point accounted for by the $r$ retained modes. Subsequently, the normalised Varimax-rotated EOFs, $\widetilde{\pmb{L}}_{V,norm}$, are the solution to the Varimax criterion,
\begin{align}\label{eq:varimax_rotation}
    \widetilde{\pmb{L}}_{V,norm} = \pmb{H}^{-1}\pmb{L}\pmb{R} \quad \text{s.t.} \quad \operatorname*{argmax}_{l} (S) \text{ and } \pmb{R}^*\pmb{R} = 1,
\end{align}
with the communality matrix $\pmb{H} \in \mathbb{R}^{n \times n}$ whose elements are given by $\text{diag}(h_1,\dots,h_n)$. Equation~\eqref{eq:varimax_rotation} can be solved by an iterative process, in which the EOFs are rotated in pairs in order to maximise $S$. Finally, the de-normalised Varimax-rotated EOFs can be computed via $\widetilde{\pmb{L}}_V = \pmb{H} \widetilde{\pmb{L}}_{V,norm}$.

\paragraph{Promax oblique rotation}
Achieving simple structures with Promax is done via an oblique Procrustes transformation \citep{hurley_procrustes_1962}. Every target matrix of rotated EOFs $\pmb{T}$ can always be approximated from a base matrix $\pmb{B}$ via a linear transformation $\pmb{R}$,
\begin{align}
    \pmb{T} = \pmb{B}\pmb{R} + \pmb{E},
\end{align}
where $\pmb{E}$ is an error matrix. Minimising $\text{trace}(\pmb{E}^*\pmb{E})$ yields the complex Procrustes equation,
\begin{align}\label{eq:procrustes}
    \pmb{R} = \left(\pmb{B}^*\pmb{B} \right)^{-1}\pmb{B}^* \pmb{T}.
\end{align}
The basic assumption of Promax is that an Varimax orthogonal rotation is a reasonable approximation to an optimal oblique solution. Therefore, the base matrix is chosen to be $\pmb{B}= \widetilde{\pmb{H}}^{-1} \widetilde{\pmb{L}}_V$ whose entries are normalised by the Varimax communalities $\widetilde{h}_{j} = \sum_{k=1}^r |\widetilde{l}_{jk}|^2$. Then, the Promax equation defines the elements of the target matrix $\pmb{T}$,
\begin{align}
    t_{jk} = |b_{jk}^+|^{p+1}/b_{jk}^+,
\end{align}
where $^+$ denotes the max-normalised entries given by $b_{jk}^+ = b_{jk} / \max_j |b_{jk}|$. The power parameter $p$ thus defines the strength of the Promax operation, while the sign remains unchanged. Using Equation~\eqref{eq:procrustes}, the de-normalised Promax-rotated EOFs are given by
\begin{align}
    \widetilde{\pmb{L}}_{P} =& \widetilde{\pmb{L}}_V \pmb{R} \pmb{D} \nonumber\\
    =& \widetilde{\pmb{H}}^{-1}\widetilde{\pmb{L}}_V \left[\left(\widetilde{\pmb{L}}_V^*\widetilde{\pmb{H}}^{-2}\widetilde{\pmb{L}}_V\right)^{-1} \widetilde{\pmb{L}}_V^*\widetilde{\pmb{H}}^{-1} \pmb{T} \right] \pmb{D},
\end{align}
where the normalisation matrix is given by $\pmb{D}^2 = \text{diag}\left(\pmb{R}^*\pmb{R}\right)^{-1}$.

%% Important!
%\appendcaption{<appendix letter and number>}{<caption>}
%must be used for figures and tables in appendixes, e.g.,
%
%\begin{figure}
%\noindent\includegraphics[width=19pc,angle=0]{figure01.pdf}\\
%\appendcaption{A1}{Caption here.}
%\end{figure}
%
% All appendix figures/tables should be placed in order AFTER the main figures/tables, i.e., tables, appendix tables, figures, appendix figures.
%
%%%%%%%%%%%%%%%%%%%%%%%%%%%%%%%%%%%%%%%%%%%%%%%%%%%%%%%%%%%%%%%%%%%%%
% REFERENCES
%%%%%%%%%%%%%%%%%%%%%%%%%%%%%%%%%%%%%%%%%%%%%%%%%%%%%%%%%%%%%%%%%%%%%
%----------------------------------------------------------------------------------------
%	BIBLIOGRAPHY
%----------------------------------------------------------------------------------------
\newpage
\printbibliography

@article{li_north_2016,
	title = {North Atlantic salinity as a predictor of Sahel rainfall},
	volume = {2},
	rights = {Copyright © 2016, The Authors. This is an open-access article distributed under the terms of the Creative Commons Attribution-{NonCommercial} license, which permits use, distribution, and reproduction in any medium, so long as the resultant use is not for commercial advantage and provided the original work is properly cited.},
	issn = {2375-2548},
	url = {https://advances.sciencemag.org/content/2/5/e1501588},
	doi = {10.1126/sciadv.1501588},
	pages = {e1501588},
	number = {5},
	journaltitle = {Science Advances},
	author = {Li, L and Schmitt, R and Ummenhofer, {CC} and Karnauskas, {KB}},
	urldate = {2019-10-28},
	date = {2016-05-01},
	langid = {english}
}

@article{bretherton_intercomparison_1992,
	title = {An intercomparison of methods for finding coupled patterns in climate data},
	volume = {5},
	pages = {541--560},
	number = {6},
	journaltitle = {Journal of climate},
	author = {Bretherton, C and Smith, C and Wallace, {JM}},
	date = {1992}
}

@article{north_sampling_1982,
	title = {Sampling Errors in the Estimation of Empirical Orthogonal Functions},
	volume = {110},
	doi = {10.1175/1520-0493(1982)110<0699:SEITEO>2.0.CO;2},
	journaltitle = {Monthly Weather Review},
	author = {North, G and L. Bell, T and Cahalan, R and J. Moeng, F},
	date = {1982}
}

@article{dee_era-interim_2011,
	title = {The {ERA}-Interim reanalysis: configuration and performance of the data assimilation system},
	volume = {137},
	issn = {0035-9009},
	url = {https://doi.org/10.1002/qj.828},
	doi = {10.1002/qj.828},
	pages = {553--597},
	number = {656},
	journaltitle = {Quarterly Journal of the Royal Meteorological Society},
	shortjournal = {Quarterly Journal of the Royal Meteorological Society},
	author = {Dee, {DP} and Uppala, {SM} and Simmons, {AJ} and Berrisford, P and Poli, P and Kobayashi, S and Andrae, U and Balmaseda, {MA} and Balsamo, G and Bauer, P and Bechtold, P and Beljaars, {ACM} and van de Berg, L and Bidlot, J and Bormann, N and Delsol, C and Dragani, R and Fuentes, M and Geer, {AJ} and Haimberger, L and Healy, {SB} and Hersbach, H and Hólm, {EV} and Isaksen, L and Kållberg, P and Köhler, M and Matricardi, M and {McNally}, {AP} and Monge-Sanz, {BM} and Morcrette, J-J and Park, B-K and Peubey, C and de Rosnay, P and Tavolato, C and Thépaut, J-N and Vitart, F},
	urldate = {2019-01-26},
	date = {2011}
}

@article{hannachi_empirical_2007,
	title = {Empirical orthogonal functions and related techniques in atmospheric science: A review},
	volume = {27},
	issn = {1097-0088},
	url = {https://rmets.onlinelibrary.wiley.com/doi/abs/10.1002/joc.1499},
	doi = {10.1002/joc.1499},
	shorttitle = {Empirical orthogonal functions and related techniques in atmospheric science},
	pages = {1119--1152},
	number = {9},
	journaltitle = {International Journal of Climatology},
	author = {Hannachi, A and Jolliffe, {IT} and Stephenson, {DB}},
	urldate = {2020-02-14},
	date = {2007},
	langid = {english}
}

@article{horel_complex_1984,
	title = {Complex Principal Component Analysis: Theory and Examples},
	volume = {23},
	issn = {0733-3021},
	url = {https://journals.ametsoc.org/doi/abs/10.1175/1520-0450(1984)023%3C1660%3ACPCATA%3E2.0.CO%3B2},
	doi = {10.1175/1520-0450(1984)023<1660:CPCATA>2.0.CO;2},
	shorttitle = {Complex Principal Component Analysis},
	pages = {1660--1673},
	number = {12},
	journaltitle = {Journal of Climate and Applied Meteorology},
	shortjournal = {J. Climate Appl. Meteor.},
	author = {Horel, {JD}},
	urldate = {2020-02-14},
	date = {1984-12-01}
}

@article{kaiser_varimax_1958,
	title = {The varimax criterion for analytic rotation in factor analysis},
	volume = {23},
	issn = {0033-3123, 1860-0980},
	url = {http://link.springer.com/10.1007/BF02289233},
	doi = {10.1007/BF02289233},
	pages = {187--200},
	number = {3},
	journaltitle = {Psychometrika},
	shortjournal = {Psychometrika},
	author = {Kaiser, Henry F.},
	urldate = {2020-04-30},
	date = {1958-09},
	langid = {english}
}

@article{hendrickson_promax_1964,
	title = {Promax: A Quick Method for Rotation to Oblique Simple Structure},
	volume = {17},
	rights = {1964 The British Psychological Society},
	issn = {2044-8317},
	url = {https://onlinelibrary.wiley.com/doi/abs/10.1111/j.2044-8317.1964.tb00244.x},
	doi = {10.1111/j.2044-8317.1964.tb00244.x},
	shorttitle = {Promax},
	pages = {65--70},
	number = {1},
	journaltitle = {British Journal of Statistical Psychology},
	author = {Hendrickson, Alan E. and White, Paul Owen},
	urldate = {2020-04-30},
	date = {1964},
	langid = {english},
	note = {\_eprint: https://onlinelibrary.wiley.com/doi/pdf/10.1111/j.2044-8317.1964.tb00244.x}
}

@article{lian_evaluation_2012,
	title = {An Evaluation of Rotated {EOF} Analysis and Its Application to Tropical Pacific {SST} Variability},
	volume = {25},
	issn = {0894-8755, 1520-0442},
	url = {http://journals.ametsoc.org/doi/10.1175/JCLI-D-11-00663.1},
	doi = {10.1175/JCLI-D-11-00663.1},
	pages = {5361--5373},
	number = {15},
	journaltitle = {Journal of Climate},
	shortjournal = {J. Climate},
	author = {Lian, Tao and Chen, Dake},
	urldate = {2020-04-30},
	date = {2012-08},
	langid = {english}
}

@article{wallace_empirical_1972,
	title = {Empirical Orthogonal Representation of Time Series in the Frequency Domain. Part I: Theoretical Considerations},
	volume = {11},
	issn = {0021-8952},
	url = {https://journals.ametsoc.org/doi/10.1175/1520-0450%281972%29011%3C0887%3AEOROTS%3E2.0.CO%3B2},
	doi = {10.1175/1520-0450(1972)011<0887:EOROTS>2.0.CO;2},
	shorttitle = {Empirical Orthogonal Representation of Time Series in the Frequency Domain. Part I},
	pages = {887--892},
	number = {6},
	journaltitle = {Journal of Applied Meteorology},
	shortjournal = {J. Appl. Meteor.},
	author = {Wallace, John M. and Dickinson, Robert E.},
	urldate = {2020-05-05},
	date = {1972-09-01},
	note = {Publisher: American Meteorological Society}
}

@article{bloomfield_orthogonal_1994,
	title = {Orthogonal rotation of complex principal components},
	volume = {14},
	rights = {Copyright © 1994 John Wiley \& Sons, Ltd},
	issn = {1097-0088},
	url = {https://rmets.onlinelibrary.wiley.com/doi/abs/10.1002/joc.3370140706},
	doi = {10.1002/joc.3370140706},
	pages = {759--775},
	number = {7},
	journaltitle = {International Journal of Climatology},
	author = {Bloomfield, Peter and Davis, Jerry M.},
	urldate = {2020-05-05},
	date = {1994},
	langid = {english},
	note = {\_eprint: https://rmets.onlinelibrary.wiley.com/doi/pdf/10.1002/joc.3370140706}
}

@article{cheng_orthogonal_1995,
	title = {Orthogonal Rotation of Spatial Patterns Derived from Singular Value Decomposition Analysis},
	volume = {8},
	issn = {0894-8755},
	url = {https://journals.ametsoc.org/jcli/article/8/11/2631/35764/Orthogonal-Rotation-of-Spatial-Patterns-Derived},
	doi = {10.1175/1520-0442(1995)008<2631:OROSPD>2.0.CO;2},
	pages = {2631--2643},
	number = {11},
	journaltitle = {Journal of Climate},
	shortjournal = {J. Climate},
	author = {Cheng, Xinhua and Dunkerton, Timothy J.},
	urldate = {2020-06-14},
	date = {1995-11-01},
	langid = {english},
	note = {Publisher: American Meteorological Society}
}

@article{ballabrerapoy_potential_2002,
	title = {On the potential impact of sea surface salinity observations on {ENSO} predictions},
	volume = {107},
	rights = {Copyright 2002 by the American Geophysical Union.},
	issn = {2156-2202},
	url = {https://agupubs.onlinelibrary.wiley.com/doi/abs/10.1029/2001JC000834},
	doi = {10.1029/2001JC000834},
	pages = {SRF 8--1--SRF 8--11},
	issue = {C12},
	journaltitle = {Journal of Geophysical Research: Oceans},
	author = {Ballabrera‐Poy, J. and Murtugudde, R. and Busalacchi, A. J.},
	urldate = {2020-06-30},
	date = {2002},
	langid = {english},
	note = {\_eprint: https://agupubs.onlinelibrary.wiley.com/doi/pdf/10.1029/2001JC000834}
}

@article{ashok_nino_2007,
	title = {El Niño Modoki and its possible teleconnection},
	volume = {112},
	rights = {Copyright 2007 by the American Geophysical Union.},
	issn = {2156-2202},
	url = {https://agupubs.onlinelibrary.wiley.com/doi/abs/10.1029/2006JC003798},
	doi = {10.1029/2006JC003798},
	issue = {C11},
	journaltitle = {Journal of Geophysical Research: Oceans},
	author = {Ashok, Karumuri and Behera, Swadhin K. and Rao, Suryachandra A. and Weng, Hengyi and Yamagata, Toshio},
	urldate = {2020-09-28},
	date = {2007},
	langid = {english},
	note = {\_eprint: https://agupubs.onlinelibrary.wiley.com/doi/pdf/10.1029/2006JC003798}
}

@article{rasmusson_biennial_1981,
	title = {Biennial variations in surface temperature over the United States as revealed by singular decomposition},
	volume = {109},
	pages = {587--598},
	number = {3},
	journaltitle = {Monthly Weather Review},
	author = {Rasmusson, Eugene M. and Arkin, Phillip A. and Chen, Wen-Yuan and Jalickee, John B.},
	date = {1981}
}

@article{hotelling_relations_1936,
	title = {Relations Between Two Sets of Variates},
	volume = {28},
	issn = {0006-3444},
	url = {https://www.jstor.org/stable/2333955},
	doi = {10.2307/2333955},
	pages = {321--377},
	number = {3},
	journaltitle = {Biometrika},
	author = {Hotelling, Harold},
	urldate = {2020-12-29},
	date = {1936},
	note = {Publisher: [Oxford University Press, Biometrika Trust]}
}

@article{feng_different_2011,
	title = {Different impacts of El Niño and El Niño Modoki on China rainfall in the decaying phases},
	volume = {31},
	issn = {08998418},
	url = {http://doi.wiley.com/10.1002/joc.2217},
	doi = {10.1002/joc.2217},
	pages = {2091--2101},
	number = {14},
	journaltitle = {International Journal of Climatology},
	shortjournal = {Int. J. Climatol.},
	author = {Feng, Juan and Chen, Wen and Tam, C.-Y. and Zhou, Wen},
	urldate = {2021-01-08},
	date = {2011-11-30},
	langid = {english}
}

@article{thurstone_simple_1947,
	title = {The simple structure concept},
	pages = {319--346},
	journaltitle = {Multiple Factor Analysis: A Development and Expansion of The Vectors of Mind},
	author = {Thurstone, L.},
	date = {1947},
	note = {Publisher: The University of Chicago Press Chicago}
}

@article{bell_era5_2020,
	title = {{ERA}5 monthly averaged data on single levels from 1950 to 1978 (preliminary version)},
	url = {https://cds.climate.copernicus.eu/cdsapp#!/dataset/reanalysis-era5-single-levels-monthly-means-preliminary-back-extension?tab=overview},
	doi = {10.24381/cds.f17050d7},
	author = {Bell, B. and Hersbach, H. and Berrisford, P and Dahlgren, P. and Horányi, A. and Muñoz Sabater, J. and Nicolas, J. and Radu, R. and Schepers, D. and Simmons, A. and Soci, C. and Thépaut, J-N.},
	urldate = {2021-01-20},
	date = {2020}
}

@article{hyndman_unmasking_2003,
	title = {Unmasking the Theta method},
	volume = {19},
	issn = {0169-2070},
	url = {http://www.sciencedirect.com/science/article/pii/S0169207001001431},
	doi = {10.1016/S0169-2070(01)00143-1},
	pages = {287--290},
	number = {2},
	journaltitle = {International Journal of Forecasting},
	shortjournal = {International Journal of Forecasting},
	author = {Hyndman, Rob J. and Billah, Baki},
	urldate = {2021-01-27},
	date = {2003-04-01},
	langid = {english}
}

@article{assimakopoulos_theta_2000,
	title = {The theta model: a decomposition approach to forecasting},
	volume = {16},
	issn = {0169-2070},
	url = {http://www.sciencedirect.com/science/article/pii/S0169207000000662},
	doi = {10.1016/S0169-2070(00)00066-2},
	series = {The M3- Competition},
	shorttitle = {The theta model},
	pages = {521--530},
	number = {4},
	journaltitle = {International Journal of Forecasting},
	shortjournal = {International Journal of Forecasting},
	author = {Assimakopoulos, V. and Nikolopoulos, K.},
	urldate = {2021-01-27},
	date = {2000-10-01},
	langid = {english}
}

@article{cheng_robustness_1995,
	title = {Robustness of Low-Frequency Circulation Patterns Derived from {EOF} and Rotated {EOF} Analyses},
	volume = {8},
	issn = {0894-8755, 1520-0442},
	url = {https://journals.ametsoc.org/view/journals/clim/8/6/1520-0442_1995_008_1709_rolfcp_2_0_co_2.xml},
	doi = {10.1175/1520-0442(1995)008<1709:ROLFCP>2.0.CO;2},
	pages = {1709--1713},
	number = {6},
	journaltitle = {Journal of Climate},
	author = {Cheng, Xinhua and Nitsche, Gregor and Wallace, John M.},
	urldate = {2021-01-28},
	date = {1995-06-01},
	note = {Publisher: American Meteorological Society
Section: Journal of Climate}
}

@article{klein_remote_1999,
	title = {Remote Sea Surface Temperature Variations during {ENSO}: Evidence for a Tropical Atmospheric Bridge},
	volume = {12},
	issn = {0894-8755, 1520-0442},
	url = {https://journals.ametsoc.org/view/journals/clim/12/4/1520-0442_1999_012_0917_rsstvd_2.0.co_2.xml},
	doi = {10.1175/1520-0442(1999)012<0917:RSSTVD>2.0.CO;2},
	shorttitle = {Remote Sea Surface Temperature Variations during {ENSO}},
	pages = {917--932},
	number = {4},
	journaltitle = {Journal of Climate},
	author = {Klein, Stephen A. and Soden, Brian J. and Lau, Ngar-Cheung},
	urldate = {2021-02-01},
	date = {1999-04-01},
	note = {Publisher: American Meteorological Society
Section: Journal of Climate}
}

@article{krishnamurthy_variability_2003,
	title = {Variability of the Indian Ocean: Relation to monsoon and {ENSO}},
	volume = {129},
	issn = {1477-870X},
	url = {https://rmets.onlinelibrary.wiley.com/doi/abs/10.1256/qj.01.166},
	doi = {https://doi.org/10.1256/qj.01.166},
	shorttitle = {Variability of the Indian Ocean},
	pages = {1623--1646},
	number = {590},
	journaltitle = {Quarterly Journal of the Royal Meteorological Society},
	author = {Krishnamurthy, V. and Kirtman, Ben P.},
	urldate = {2021-02-01},
	date = {2003},
	langid = {english},
	note = {\_eprint: https://rmets.onlinelibrary.wiley.com/doi/pdf/10.1256/qj.01.166}
}

@article{dai_global_2000,
	title = {Global patterns of {ENSO}-induced precipitation},
	volume = {27},
	issn = {1944-8007},
	url = {https://agupubs.onlinelibrary.wiley.com/doi/abs/10.1029/1999GL011140},
	doi = {https://doi.org/10.1029/1999GL011140},
	pages = {1283--1286},
	number = {9},
	journaltitle = {Geophysical Research Letters},
	author = {Dai, Aiguo and Wigley, T. M. L.},
	urldate = {2021-02-04},
	date = {2000},
	langid = {english},
	note = {\_eprint: https://agupubs.onlinelibrary.wiley.com/doi/pdf/10.1029/1999GL011140}
}

@article{ropelewski_north_1986,
	title = {North American Precipitation and Temperature Patterns Associated with the El Niño/Southern Oscillation ({ENSO})},
	volume = {114},
	issn = {1520-0493, 0027-0644},
	url = {https://journals.ametsoc.org/view/journals/mwre/114/12/1520-0493_1986_114_2352_napatp_2_0_co_2.xml},
	doi = {10.1175/1520-0493(1986)114<2352:NAPATP>2.0.CO;2},
	pages = {2352--2362},
	number = {12},
	journaltitle = {Monthly Weather Review},
	author = {Ropelewski, C. F. and Halpert, M. S.},
	urldate = {2021-02-04},
	date = {1986-12-01},
	note = {Publisher: American Meteorological Society
Section: Monthly Weather Review}
}

@article{indeje_enso_2000,
	title = {{ENSO} signals in East African rainfall seasons},
	volume = {20},
	pages = {19--46},
	number = {1},
	journaltitle = {International Journal of Climatology: A Journal of the Royal Meteorological Society},
	author = {Indeje, Matayo and Semazzi, Fredrick {HM} and Ogallo, Laban J.},
	date = {2000},
	note = {Publisher: Wiley Online Library}
}

@article{gaughan_inter-_2016,
	title = {Inter- and Intra-annual precipitation variability and associated relationships to {ENSO} and the {IOD} in southern Africa},
	volume = {36},
	rights = {© 2015 Royal Meteorological Society},
	issn = {1097-0088},
	url = {https://rmets.onlinelibrary.wiley.com/doi/abs/10.1002/joc.4448},
	doi = {https://doi.org/10.1002/joc.4448},
	pages = {1643--1656},
	number = {4},
	journaltitle = {International Journal of Climatology},
	author = {Gaughan, Andrea E. and Staub, Caroline G. and Hoell, Andrew and Weaver, Ariel and Waylen, Peter R.},
	urldate = {2021-02-04},
	date = {2016},
	langid = {english},
	note = {\_eprint: https://rmets.onlinelibrary.wiley.com/doi/pdf/10.1002/joc.4448}
}

@article{wen_direct_2019,
	title = {Direct {ENSO} impact on East Asian summer precipitation in the developing summer},
	volume = {52},
	issn = {1432-0894},
	url = {https://doi.org/10.1007/s00382-018-4545-0},
	doi = {10.1007/s00382-018-4545-0},
	pages = {6799--6815},
	number = {11},
	journaltitle = {Climate Dynamics},
	shortjournal = {Clim Dyn},
	author = {Wen, Na and Liu, Zhengyu and Li, Laurent},
	urldate = {2021-02-04},
	date = {2019-06-01},
	langid = {english}
}

@article{cherchi_influence_2013,
	title = {Influence of {ENSO} and of the Indian Ocean Dipole on the Indian summer monsoon variability},
	volume = {41},
	pages = {81--103},
	number = {1},
	journaltitle = {Climate dynamics},
	author = {Cherchi, Annalisa and Navarra, Antonio},
	date = {2013},
	note = {Publisher: Springer}
}

@article{taschetto_nino_2009,
	title = {El Niño Modoki impacts on Australian rainfall},
	volume = {22},
	pages = {3167--3174},
	number = {11},
	journaltitle = {Journal of Climate},
	author = {Taschetto, Andréa S. and England, Matthew H.},
	date = {2009}
}

@article{ratnam_remote_2014,
	title = {Remote effects of El Niño and Modoki events on the austral summer precipitation of southern Africa},
	volume = {27},
	pages = {3802--3815},
	number = {10},
	journaltitle = {Journal of Climate},
	author = {Ratnam, J. V. and Behera, S. K. and Masumoto, Y. and Yamagata, T.},
	date = {2014}
}

@article{tedeschi_influences_2013,
	title = {Influences of two types of {ENSO} on South American precipitation},
	volume = {33},
	pages = {1382--1400},
	number = {6},
	journaltitle = {International Journal of Climatology},
	author = {Tedeschi, Renata G. and Cavalcanti, Iracema {FA} and Grimm, Alice M.},
	date = {2013},
	note = {Publisher: Wiley Online Library}
}

@report{simmons_low_2021,
	title = {Low frequency variability and trends in surface air temperature and humidity from {ERA}5 and other datasets},
	url = {https://www.ecmwf.int/sites/default/files/elibrary/2021/19911-low-frequency-variability-and-trends-surface-air-temperature-and-humidity-era5-and-other.pdf},
	shorttitle = {{ECMWF} Technical Memoranda},
	number = {881},
	institution = {{ECMWF}},
	type = {{ECMWF} Technical Memoranda},
	author = {Simmons, Adrian and Hersbach, Hans and Munoz-Sabater, Joaquin and Nicolas, Julien and Vamborg, Freja and Berrisford, Paul and de Rosnay, Patricia and Willett, Kate and Woollen, Jack},
	urldate = {2021-08-02},
	date = {2021-02},
	doi = {10.21957/ly5vbtbfd}
}

@article{boashash_estimating_1992,
	title = {Estimating and interpreting the instantaneous frequency of a signal. I. Fundamentals},
	volume = {80},
	issn = {1558-2256},
	doi = {10.1109/5.135376},
	pages = {520--538},
	number = {4},
	journaltitle = {Proceedings of the {IEEE}},
	author = {Boashash, B.},
	date = {1992-04},
	note = {Conference Name: Proceedings of the {IEEE}}
}

@article{finch_comparison_2006,
	title = {Comparison of the Performance of Varimax and Promax Rotations: Factor Structure Recovery for Dichotomous Items},
	volume = {43},
	issn = {1745-3984},
	url = {https://onlinelibrary.wiley.com/doi/abs/10.1111/j.1745-3984.2006.00003.x},
	doi = {https://doi.org/10.1111/j.1745-3984.2006.00003.x},
	shorttitle = {Comparison of the Performance of Varimax and Promax Rotations},
	pages = {39--52},
	number = {1},
	journaltitle = {Journal of Educational Measurement},
	author = {Finch, Holmes},
	urldate = {2021-02-13},
	date = {2006},
	langid = {english},
	note = {\_eprint: https://onlinelibrary.wiley.com/doi/pdf/10.1111/j.1745-3984.2006.00003.x}
}

@article{fiorucci_models_2016,
	title = {Models for optimising the theta method and their relationship to state space models},
	volume = {32},
	issn = {01692070},
	url = {https://linkinghub.elsevier.com/retrieve/pii/S0169207016300243},
	doi = {10.1016/j.ijforecast.2016.02.005},
	pages = {1151--1161},
	number = {4},
	journaltitle = {International Journal of Forecasting},
	shortjournal = {International Journal of Forecasting},
	author = {Fiorucci, Jose A. and Pellegrini, Tiago R. and Louzada, Francisco and Petropoulos, Fotios and Koehler, Anne B.},
	urldate = {2021-02-13},
	date = {2016-10},
	langid = {english}
}

@article{wiedermann_differential_2021,
	title = {Differential Imprints of Distinct {ENSO} Flavors in Global Patterns of Very Low and High Seasonal Precipitation},
	volume = {3},
	issn = {2624-9553},
	url = {https://www.frontiersin.org/articles/10.3389/fclim.2021.618548/full},
	doi = {10.3389/fclim.2021.618548},
	journaltitle = {Frontiers in Climate},
	shortjournal = {Front. Clim.},
	author = {Wiedermann, Marc and Siegmund, Jonatan F. and Donges, Jonathan F. and Donner, Reik V.},
	urldate = {2021-02-16},
	date = {2021},
	note = {Publisher: Frontiers}
}

@article{xie_global_2010,
	title = {Global Warming Pattern Formation: Sea Surface Temperature and Rainfall},
	volume = {23},
	issn = {0894-8755, 1520-0442},
	url = {https://journals.ametsoc.org/view/journals/clim/23/4/2009jcli3329.1.xml},
	doi = {10.1175/2009JCLI3329.1},
	shorttitle = {Global Warming Pattern Formation},
	pages = {966--986},
	number = {4},
	journaltitle = {Journal of Climate},
	author = {Xie, Shang-Ping and Deser, Clara and Vecchi, Gabriel A. and Ma, Jian and Teng, Haiyan and Wittenberg, Andrew T.},
	urldate = {2021-02-25},
	date = {2010-02-15},
	note = {Publisher: American Meteorological Society
Section: Journal of Climate}
}

@article{giorgi_response_2019,
	title = {The response of precipitation characteristics to global warming from climate projections},
	volume = {10},
	issn = {2190-4979},
	url = {https://esd.copernicus.org/articles/10/73/2019/},
	doi = {https://doi.org/10.5194/esd-10-73-2019},
	pages = {73--89},
	number = {1},
	journaltitle = {Earth System Dynamics},
	author = {Giorgi, Filippo and Raffaele, Francesca and Coppola, Erika},
	urldate = {2021-02-25},
	date = {2019-02-06},
	note = {Publisher: Copernicus {GmbH}}
}

@article{gu_spatial_2015,
	title = {Spatial Patterns of Global Precipitation Change and Variability during 1901–2010},
	volume = {28},
	issn = {0894-8755, 1520-0442},
	url = {https://journals.ametsoc.org/view/journals/clim/28/11/jcli-d-14-00201.1.xml},
	doi = {10.1175/JCLI-D-14-00201.1},
	pages = {4431--4453},
	number = {11},
	journaltitle = {Journal of Climate},
	author = {Gu, Guojun and Adler, Robert F.},
	urldate = {2021-02-25},
	date = {2015-06-01},
	note = {Publisher: American Meteorological Society
Section: Journal of Climate}
}

@article{newman_pacific_2016,
	title = {The Pacific Decadal Oscillation, Revisited},
	volume = {29},
	issn = {0894-8755, 1520-0442},
	url = {https://journals.ametsoc.org/view/journals/clim/29/12/jcli-d-15-0508.1.xml},
	doi = {10.1175/JCLI-D-15-0508.1},
	pages = {4399--4427},
	number = {12},
	journaltitle = {Journal of Climate},
	author = {Newman, Matthew and Alexander, Michael A. and Ault, Toby R. and Cobb, Kim M. and Deser, Clara and Lorenzo, Emanuele Di and Mantua, Nathan J. and Miller, Arthur J. and Minobe, Shoshiro and Nakamura, Hisashi and Schneider, Niklas and Vimont, Daniel J. and Phillips, Adam S. and Scott, James D. and Smith, Catherine A.},
	urldate = {2021-02-26},
	date = {2016-06-15},
	note = {Publisher: American Meteorological Society
Section: Journal of Climate}
}

@article{mantua_pacific_1997,
	title = {A Pacific Interdecadal Climate Oscillation with Impacts on Salmon Production*},
	volume = {78},
	issn = {0003-0007, 1520-0477},
	url = {https://journals.ametsoc.org/view/journals/bams/78/6/1520-0477_1997_078_1069_apicow_2_0_co_2.xml},
	doi = {10.1175/1520-0477(1997)078<1069:APICOW>2.0.CO;2},
	pages = {1069--1080},
	number = {6},
	journaltitle = {Bulletin of the American Meteorological Society},
	author = {Mantua, Nathan J. and Hare, Steven R. and Zhang, Yuan and Wallace, John M. and Francis, Robert C.},
	urldate = {2021-02-26},
	date = {1997-06-01},
	langid = {english},
	note = {Publisher: American Meteorological Society
Section: Bulletin of the American Meteorological Society}
}

@article{pierce_role_2002,
	title = {The Role of Sea Surface Temperatures in Interactions between {ENSO} and the North Pacific Oscillation},
	volume = {15},
	issn = {0894-8755, 1520-0442},
	url = {https://journals.ametsoc.org/view/journals/clim/15/11/1520-0442_2002_015_1295_trosst_2.0.co_2.xml},
	doi = {10.1175/1520-0442(2002)015<1295:TROSST>2.0.CO;2},
	pages = {1295--1308},
	number = {11},
	journaltitle = {Journal of Climate},
	author = {Pierce, David W.},
	urldate = {2021-02-26},
	date = {2002-06-01},
	note = {Publisher: American Meteorological Society
Section: Journal of Climate}
}

@article{servain_relationship_1999,
	title = {Relationship between the equatorial and meridional modes of climatic variability in the tropical Atlantic},
	volume = {26},
	rights = {Copyright 1999 by the American Geophysical Union.},
	issn = {1944-8007},
	url = {https://agupubs.onlinelibrary.wiley.com/doi/abs/10.1029/1999GL900014},
	doi = {https://doi.org/10.1029/1999GL900014},
	pages = {485--488},
	number = {4},
	journaltitle = {Geophysical Research Letters},
	author = {Servain, Jacques and Wainer, Ilana and {McCreary}, Julian P. and Dessier, Alain},
	urldate = {2021-02-27},
	date = {1999},
	langid = {english},
	note = {\_eprint: https://agupubs.onlinelibrary.wiley.com/doi/pdf/10.1029/1999GL900014}
}

@article{chiang_analogous_2004,
	title = {Analogous Pacific and Atlantic Meridional Modes of Tropical Atmosphere–Ocean Variability},
	volume = {17},
	issn = {0894-8755, 1520-0442},
	url = {https://journals.ametsoc.org/view/journals/clim/17/21/jcli4953.1.xml},
	doi = {10.1175/JCLI4953.1},
	pages = {4143--4158},
	number = {21},
	journaltitle = {Journal of Climate},
	author = {Chiang, John C. H. and Vimont, Daniel J.},
	urldate = {2021-02-27},
	date = {2004-11-01},
	note = {Publisher: American Meteorological Society
Section: Journal of Climate}
}

@article{martin_multidecadal_2014,
	title = {The Multidecadal Atlantic {SST}—Sahel Rainfall Teleconnection in {CMIP}5 Simulations},
	volume = {27},
	issn = {0894-8755, 1520-0442},
	url = {https://journals.ametsoc.org/view/journals/clim/27/2/jcli-d-13-00242.1.xml},
	doi = {10.1175/JCLI-D-13-00242.1},
	pages = {784--806},
	number = {2},
	journaltitle = {Journal of Climate},
	author = {Martin, Elinor R. and Thorncroft, Chris and Booth, Ben B. B.},
	urldate = {2021-02-27},
	date = {2014-01-15},
	note = {Publisher: American Meteorological Society
Section: Journal of Climate}
}

@article{lamb_interannual_1986,
	title = {Interannual variability in the tropical Atlantic},
	volume = {322},
	rights = {1986 Nature Publishing Group},
	issn = {1476-4687},
	url = {https://www.nature.com/articles/322238a0},
	doi = {10.1038/322238a0},
	pages = {238--240},
	number = {6076},
	journaltitle = {Nature},
	author = {Lamb, Peter J. and Peppler, Randy A. and Hastenrath, Stefan},
	urldate = {2021-02-27},
	date = {1986-07},
	langid = {english},
	note = {Number: 6076
Publisher: Nature Publishing Group}
}

@article{vittal_early_2020,
	title = {Early prediction of the Indian summer monsoon rainfall by the Atlantic Meridional Mode},
	volume = {54},
	issn = {1432-0894},
	url = {https://doi.org/10.1007/s00382-019-05117-0},
	doi = {10.1007/s00382-019-05117-0},
	pages = {2337--2346},
	number = {3},
	journaltitle = {Climate Dynamics},
	shortjournal = {Clim Dyn},
	author = {Vittal, H. and Villarini, Gabriele and Zhang, Wei},
	urldate = {2021-02-27},
	date = {2020-02-01},
	langid = {english}
}

@article{kutzbach_empirical_1967,
	title = {Empirical eigenvectors of sea-level pressure, surface temperature and precipitation complexes over North America},
	volume = {6},
	pages = {791--802},
	number = {5},
	journaltitle = {Journal of Applied Meteorology and Climatology},
	author = {Kutzbach, John E.},
	date = {1967}
}

@article{vinod_canonical_1976,
	title = {Canonical ridge and econometrics of joint production},
	volume = {4},
	pages = {147--166},
	number = {2},
	journaltitle = {Journal of econometrics},
	author = {Vinod, Hrishikesh D.},
	date = {1976},
	note = {Publisher: Elsevier}
}

@article{cruz-cano_fast_2014,
	title = {Fast regularized canonical correlation analysis},
	volume = {70},
	issn = {0167-9473},
	url = {https://www.sciencedirect.com/science/article/pii/S0167947313003447},
	doi = {10.1016/j.csda.2013.09.020},
	pages = {88--100},
	journaltitle = {Computational Statistics \& Data Analysis},
	shortjournal = {Computational Statistics \& Data Analysis},
	author = {Cruz-Cano, Raul and Lee, Mei-Ling Ting},
	urldate = {2021-02-27},
	date = {2014-02-01},
	langid = {english}
}

@article{mantua_pacific_2002,
	title = {The Pacific decadal oscillation},
	volume = {58},
	pages = {35--44},
	number = {1},
	journaltitle = {Journal of oceanography},
	author = {Mantua, Nathan J. and Hare, Steven R.},
	date = {2002},
	note = {Publisher: Springer}
}

@article{enfield_tropical_1997,
	title = {Tropical Atlantic sea surface temperature variability and its relation to El Niño-Southern Oscillation},
	volume = {102},
	rights = {This paper is not subject to U.S. copyright. Published in 1997 by the American Geophysical Union.},
	issn = {2156-2202},
	url = {https://agupubs.onlinelibrary.wiley.com/doi/abs/10.1029/96JC03296},
	doi = {https://doi.org/10.1029/96JC03296},
	pages = {929--945},
	issue = {C1},
	journaltitle = {Journal of Geophysical Research: Oceans},
	author = {Enfield, David B. and Mayer, Dennis A.},
	urldate = {2021-03-12},
	date = {1997},
	langid = {english},
	note = {\_eprint: https://agupubs.onlinelibrary.wiley.com/doi/pdf/10.1029/96JC03296}
}

@article{saravanan_interaction_2000,
	title = {Interaction between Tropical Atlantic Variability and El Niño–Southern Oscillation},
	volume = {13},
	issn = {0894-8755, 1520-0442},
	url = {https://journals.ametsoc.org/view/journals/clim/13/13/1520-0442_2000_013_2177_ibtava_2.0.co_2.xml},
	doi = {10.1175/1520-0442(2000)013<2177:IBTAVA>2.0.CO;2},
	pages = {2177--2194},
	number = {13},
	journaltitle = {Journal of Climate},
	author = {Saravanan, R. and Chang, Ping},
	urldate = {2021-03-12},
	date = {2000-07-01},
	note = {Publisher: American Meteorological Society
Section: Journal of Climate}
}

@article{alexander_influence_2002,
	title = {The influence of {ENSO} on air-sea interaction in the Atlantic},
	volume = {29},
	rights = {Copyright 2002 by the American Geophysical Union.},
	issn = {1944-8007},
	url = {https://agupubs.onlinelibrary.wiley.com/doi/abs/10.1029/2001GL014347},
	doi = {https://doi.org/10.1029/2001GL014347},
	pages = {46--1--46--4},
	number = {14},
	journaltitle = {Geophysical Research Letters},
	author = {Alexander, Michael and Scott, James},
	urldate = {2021-03-12},
	date = {2002},
	langid = {english},
	note = {\_eprint: https://agupubs.onlinelibrary.wiley.com/doi/pdf/10.1029/2001GL014347}
}

@article{chiang_tropical_2002,
	title = {Tropical Tropospheric Temperature Variations Caused by {ENSO} and Their Influence on the Remote Tropical Climate},
	volume = {15},
	issn = {0894-8755, 1520-0442},
	url = {https://journals.ametsoc.org/view/journals/clim/15/18/1520-0442_2002_015_2616_tttvcb_2.0.co_2.xml},
	doi = {10.1175/1520-0442(2002)015<2616:TTTVCB>2.0.CO;2},
	pages = {2616--2631},
	number = {18},
	journaltitle = {Journal of Climate},
	author = {Chiang, John C. H. and Sobel, Adam H.},
	urldate = {2021-03-12},
	date = {2002-09-15},
	note = {Publisher: American Meteorological Society
Section: Journal of Climate}
}

@article{lau_role_1996,
	title = {The role of the “atmospheric bridge” in linking tropical Pacific {ENSO} events to extratropical {SST} anomalies},
	volume = {9},
	pages = {2036--2057},
	number = {9},
	journaltitle = {Journal of Climate},
	author = {Lau, Ngar-Cheung and Nath, Mary Jo},
	date = {1996}
}

@article{hurley_procrustes_1962,
	title = {The Procrustes program: Producing direct rotation to test a hypothesized factor structure},
	volume = {7},
	shorttitle = {The Procrustes program},
	pages = {258},
	number = {2},
	journaltitle = {Behavioral science},
	author = {Hurley, John R. and Cattell, Raymond B.},
	date = {1962},
	note = {Publisher: University of Michigan, Mental Health Research Institute}
}

@article{hersbach_era5_2019,
	title = {{ERA}5 monthly averaged data on single levels from 1979 to present},
	doi = {10.24381/cds.f17050d7},
	author = {Hersbach, H. and Bell, B. and Berrisford, P and Biavati, G. and Horányi, A. and Muñoz Sabater, J. and Nicolas, J. and Peubey, C. and Radu, R. and Rozum, I. and Schepers, D. and Simmons, A. and Soci, C. and Dee, D. and Thépaut, J-N.},
	urldate = {2021-01-20},
	date = {2019}
}

@article{lenssen_seasonal_2020,
	title = {Seasonal Forecast Skill of {ENSO} Teleconnection Maps},
	volume = {35},
	issn = {1520-0434, 0882-8156},
	url = {https://journals.ametsoc.org/view/journals/wefo/35/6/WAF-D-19-0235.1.xml},
	doi = {10.1175/WAF-D-19-0235.1},
	pages = {2387--2406},
	number = {6},
	journaltitle = {Weather and Forecasting},
	author = {Lenssen, Nathan J. L. and Goddard, Lisa and Mason, Simon},
	urldate = {2021-03-22},
	date = {2020-12-01},
	note = {Publisher: American Meteorological Society
Section: Weather and Forecasting}
}

@article{zhang_madden-julian_2005,
	title = {Madden-Julian Oscillation},
	volume = {43},
	issn = {1944-9208},
	url = {https://agupubs.onlinelibrary.wiley.com/doi/abs/10.1029/2004RG000158},
	doi = {10.1029/2004RG000158},
	number = {2},
	journaltitle = {Reviews of Geophysics},
	author = {Zhang, Chidong},
	urldate = {2021-07-24},
	date = {2005},
	langid = {english},
	note = {eprint: https://agupubs.onlinelibrary.wiley.com/doi/pdf/10.1029/2004RG000158}
}

@article{madden_detection_1971,
	title = {Detection of a 40–50 Day Oscillation in the Zonal Wind in the Tropical Pacific},
	volume = {28},
	issn = {0022-4928, 1520-0469},
	url = {https://journals.ametsoc.org/view/journals/atsc/28/5/1520-0469_1971_028_0702_doadoi_2_0_co_2.xml},
	doi = {10.1175/1520-0469(1971)028<0702:DOADOI>2.0.CO;2},
	pages = {702--708},
	number = {5},
	journaltitle = {Journal of the Atmospheric Sciences},
	author = {Madden, Roland A. and Julian, Paul R.},
	urldate = {2021-07-24},
	date = {1971-07-01},
	note = {Publisher: American Meteorological Society
Section: Journal of the Atmospheric Sciences}
}

@article{richman_rotation_1986,
	title = {Rotation of principal components},
	volume = {6},
	pages = {293--335},
	number = {3},
	journaltitle = {Journal of climatology},
	author = {Richman, Michael B.},
	date = {1986},
	note = {Publisher: Wiley Online Library}
}

@article{hu_interferential_2009,
	title = {Interferential Impact of {ENSO} and {PDO} on Dry and Wet Conditions in the U.S. Great Plains},
	volume = {22},
	issn = {0894-8755, 1520-0442},
	url = {https://journals.ametsoc.org/view/journals/clim/22/22/2009jcli2798.1.xml},
	doi = {10.1175/2009JCLI2798.1},
	pages = {6047--6065},
	number = {22},
	journaltitle = {Journal of Climate},
	author = {Hu, Zeng-Zhen and Huang, Bohua},
	urldate = {2021-07-30},
	date = {2009-11-15},
	note = {Publisher: American Meteorological Society
Section: Journal of Climate}
}

@article{hannachi_regularised_2016,
	title = {Regularised empirical orthogonal functions},
	volume = {68},
	pages = {31723},
	number = {1},
	journaltitle = {Tellus A: Dynamic Meteorology and Oceanography},
	author = {Hannachi, Abdel},
	date = {2016},
	note = {Publisher: Taylor \& Francis}
}

@book{hannachi_patterns_2021,
	title = {Patterns Identification and Data Mining in Weather and Climate},
	isbn = {978-3-030-67072-6},
	series = {Springer Atmospheric Sciences},
	pagetotal = {600},
	publisher = {Springer International Publishing},
	author = {Hannachi, Abdelwaheb},
	date = {2021}
}

@article{hannachi_archetypal_2017,
	title = {Archetypal Analysis: Mining Weather and Climate Extremes},
	volume = {30},
	issn = {0894-8755, 1520-0442},
	url = {https://journals.ametsoc.org/view/journals/clim/30/17/jcli-d-16-0798.1.xml},
	doi = {10.1175/JCLI-D-16-0798.1},
	shorttitle = {Archetypal Analysis},
	pages = {6927--6944},
	number = {17},
	journaltitle = {Journal of Climate},
	author = {Hannachi, A. and Trendafilov, N.},
	urldate = {2021-08-02},
	date = {2017-09-01},
	note = {Publisher: American Meteorological Society
Section: Journal of Climate}
}

@article{rieger_xmca_2021,
	title = {xmca v0.3.3: Maximum Covariance Analysis for Climate Science},
	rights = {{MIT}},
	url = {https://github.com/nicrie/xmca/tree/0.3.3},
	doi = {10.5281/zenodo.4751735},
	shorttitle = {xmca},
	author = {Rieger, Niclas},
	date = {2021-05}
}

@article{van_der_walt_numpy_2011,
	title = {The {NumPy} array: a structure for efficient numerical computation},
	volume = {13},
	shorttitle = {The {NumPy} array},
	pages = {22--30},
	number = {2},
	journaltitle = {Computing in science \& engineering},
	author = {Van Der Walt, Stefan and Colbert, S. Chris and Varoquaux, Gael},
	date = {2011},
	note = {Publisher: {IEEE}}
}

@article{hoyer_xarray_2017,
	title = {xarray: N-D labeled Arrays and Datasets in Python},
	volume = {5},
	rights = {Authors who publish with this journal agree to the following terms:    Authors retain copyright and grant the journal right of first publication with the work simultaneously licensed under a  Creative Commons Attribution License  that allows others to share the work with an acknowledgement of the work's authorship and initial publication in this journal.  Authors are able to enter into separate, additional contractual arrangements for the non-exclusive distribution of the journal's published version of the work (e.g., post it to an institutional repository or publish it in a book), with an acknowledgement of its initial publication in this journal.  Authors are permitted and encouraged to post their work online (e.g., in institutional repositories or on their website) prior to and during the submission process, as it can lead to productive exchanges, as well as earlier and greater citation of published work (See  The Effect of Open Access ).  All third-party images reproduced on this journal are shared under Educational Fair Use. For more information on  Educational Fair Use , please see  this useful checklist prepared by Columbia University Libraries .   All copyright  of third-party content posted here for research purposes belongs to its original owners.  Unless otherwise stated all references to characters and comic art presented on this journal are ©, ® or ™ of their respective owners. No challenge to any owner’s rights is intended or should be inferred.},
	issn = {2049-9647},
	url = {http://openresearchsoftware.metajnl.com/articles/10.5334/jors.148/},
	doi = {10.5334/jors.148},
	shorttitle = {xarray},
	pages = {10},
	number = {1},
	journaltitle = {Journal of Open Research Software},
	author = {Hoyer, Stephan and Hamman, Joe},
	urldate = {2021-08-06},
	date = {2017-04-05},
	langid = {english},
	note = {Number: 1
Publisher: Ubiquity Press}
}

\end{document}